\documentclass{aa}  

\usepackage{graphicx}
\usepackage{subfigure}
\usepackage{amssymb}
\usepackage[utf8]{inputenc}
\usepackage{amsmath}
\usepackage{lineno}
\usepackage[squaren]{SIunits}
\usepackage{nicefrac}
\usepackage{placeins}
\usepackage[version=4]{mhchem}
\usepackage{soul}
\usepackage{multirow}
\usepackage{pdfpages}
\usepackage{lineno}
\usepackage{hyperref}
\hypersetup{colorlinks=true, linkcolor=blue, breaklinks=true,  urlcolor= blue, citecolor=blue} 
\usepackage[varg]{txfonts}

\usepackage[normalem]{ulem}
\usepackage{color}

\begin{document}

   \title{The surface of (4) Vesta in visible light as seen by Dawn/{VIR}}
   \subtitle{Reproduced with permission from Astronomy \& Astrophysics, \copyright ESO}

\author{B. Rousseau \inst{1}
  \and M.C. De Sanctis \inst{1}
  \and A. Raponi \inst{1}
  \and M. Ciarniello \inst{1}
  \and E. Ammannito \inst{2}
  \and A. Frigeri \inst{1}
  \and \mbox{F.G. Carrozzo \inst{1}}
  \and \mbox{F. Tosi \inst{1}}
  \and \mbox{P. Scarica \inst{1}}
  \and S. Fonte \inst{1}
  \and C. A. Raymond \inst{3}
  \and C. T. Russell \inst{4}
  }

   \institute{Istituto Nazionale di Astrofisica (INAF) - Istituto di Astrofisica e Planetologia Spaziali (IAPS), Via Fosso del Cavaliere, 100, 00133, Rome, Italy\\
              \email{batiste.rousseau@inaf.it}
         \and
             Italian Space Agency (ASI), Via del Politecnico, 00133, Rome, Italy
         \and
             Jet Propulsion Laboratory, California Institute of Technology, Pasadena, USA
         \and
             University of California Los Angeles, Earth Planetary and Space Sciences, Los Angeles, CA, USA
            }

   \date{Received 9 June 2021; accepted 12 August 2021}

  
  \abstract
   {}
   {We analyzed the surface of Vesta at visible wavelengths, using the data of the Visible and InfraRed mapping spectrometer (VIR) on board the Dawn spacecraft. We mapped the variations of various spectral parameters on the entire surface of the asteroid, and  also derived a map of the lithology.}
   {We took advantage of the recent corrected VIR visible data to map the radiance factor at \unit{550}{\nano\meter}, three color composites, two spectral slopes, and a band area parameter relative to the \unit{930}{\nano\meter} crystal field signature in pyroxene. Using the howardite-eucrite-diogenite (HED) meteorites data as a reference, we derived the lithology of Vesta using the variations of the \unit{930}{\nano\meter} and \unit{506}{\nano\meter} (spin-forbidden) band centers observed in the VIR dataset.}
   {Our spectral parameters highlight a significant spectral diversity at the surface of Vesta. This diversity is mainly evidenced by impact craters and illustrates the heterogeneous subsurface and upper crust of Vesta. Impact craters also participate directly in this spectral diversity by bringing dark exogenous material to an almost entire hemisphere. Our derived lithology agrees with previous results obtained using a combination of infrared and visible data. We therefore demonstrate that it is possible to obtain crucial mineralogical information from visible wavelengths alone. In addition to the \unit{506}{\nano\meter} band, we identified the \unit{550}{\nano\meter} spin-forbidden one. As reported by a laboratory study for synthetic pyroxenes, we also do not observe any shift of the band center of this feature across the surface of Vesta, and thus across different mineralogies, preventing use of the \unit{550}{\nano\meter} spin-forbidden band for the lithology derivation. Finally, the largest previously identified olivine rich-spot shows a peculiar behavior in two color composites but not in the other spectral parameters.}
   {}


   \maketitle
%
\section{Introduction}
\label{Sec_Introduction}
The asteroid (4) Vesta represents about $9\%$ of the asteroid belt mass of which it is the second most massive body after the dwarf planet (1) Ceres \citep{2000_Michalak,2011_Konopliv_a,2012_Russell}. Early spectroscopic observations of Vesta by \cite{1970_McCord} showed a basaltic asteroid that is dominated by pyroxene signatures because of the crystal field electronic transition of \ce{Fe2+} within the pyroxene M2 site \citep{1970_Burns}. For this reason, Vesta was already the principal candidate to be the parent body of the howardite-eucrite-diogenite (HED) clan of meteorites \citep{1970_McCord}. Various spectroscopic, petrologic, and dynamic studies strengthened this hypothesis (\cite{1977_Consolmagno,1980_Feierberg,1985_Wisdom,1997_Binzel,1997_Gaffey} and references therein). The discovery of the V-type spectral class asteroids belonging to the Vesta asteroidal family (the Vestoids, \cite{1993_Binzel_b}), which would have been ejected from Vesta after a giant impact, reinforced the hypothesis that Vesta was the parent body of the HED meteorites. The observation of a large impact basin that marks Vesta's south pole \citep{1997_Thomas} demonstrated the collisional origin of the Vesta family. The Vestoids are thought to be the immediate parent objects for most HED meteorites because of the roles of the resonances that act as escape hatches from the Main Belt and provide trajectories into the inner Solar System.\par
Vesta can be seen as a true differentiated proto-planet. HED studies indicate that diogenites are mainly constituted of Mg-rich orthopyroxene, while eucrites (basaltic or cumulate-type) are made of pyroxene and plagioclase; although these latter span a wide range of mineralogy and textures \cite{2011_McSween}. Howardites are breccias composed of diogenites and eucrites. Considering an evolutionary model based on an initial magma ocean (e.g. \cite{1997_Righter}), the mantle would be made of olivine, the lower crust of diogenite, and the upper crust of eucrite. Nonetheless, the analysis of some HEDs, and later Vesta's gravity anomalies, indicates that alternative evolutionary paths are possible and that implementations of diogenitic plutons in the crust cannot be ruled out \citep{1995_Fowler,2017_Raymond}.\par
At the dawn of the Dawn mission, the understanding of the HEDs and their relation to Vesta led to a consistent view of the history of this latter, with a solid footing \citep{2011_McSween,2011_Coradini_a,2011_Russell}. The NASA space mission was designed to answer the remaining questions and between July 2011 and September 2012 the Dawn spacecraft took advantage of a close-up view of Vesta (from \unit{3000}{\kilo\meter} to \unit{265}{\kilo\meter}) to study its surface through three instruments: the Framing Camera (FC, \cite{2011_Sierks}), the Gamma Ray and Neutron Detector (GRaND, \cite{2011_Prettyman}), and the Visible and InfraRed spectrometer (VIR, \cite{2011_De_Sanctis_a}). Radio science was also used to study the Vesta internal structure \citep{2011_Konopliv_b,2014_Konopliv}. Dawn observed the huge impact that formed the basin on Vesta's south pole, now named Rheasilvia, which is thought to be the origin of the Vestoids. Thanks to these observations, the longitudinal and latitudinal dichotomies previously observed by the Hubble Space Telescope \citep{1997_Binzel,2010_Li} have been drastically resolved \citep{2012_Reddy_a,2012_De_Sanctis_a,2013_Ammannito_a}. The VIR spectrometer also confirmed the ubiquitous presence of pyroxene signatures \citep{2012_De_Sanctis_a} around \unit{0.9}{\micro\meter} and \unit{1.9}{\micro\meter} (hereafter BI and BII respectively), as previously observed \citep{1970_McCord,1997_Gaffey}. The weak \unit{506}{\nano\meter}-absorption (hereafter SF1), which was reported through ground-based and Hubble Space Telescope observations \citep{1997_Golubeva_Shestopalov, 1998_Cochran_Vilas}, has also been partially mapped with VIR data by \cite{2015_Stephan}. The SF1 band and the \unit{550}{\nano\meter} band (hereafter SF2) are also observed in several Vestoid spectra (\cite{2000_Vilas,2008_Shestopalov, 2021_Migliorini}, and references therein). These bands originate from the \ce{Fe2+} spin-forbidden transition in the pyroxene M1 (SF1) and M2 (SF2) sites \citep{1970_Burns,1986_SangBo,2007_Klima}.\par
Studies of the distribution of the pyroxene bands on the surface of  Vesta show mainly an eucritic and howarditic composition, with only a few localized diogenite signatures generally corresponding to the floor and the rims of the Rheasilvia basin \citep{2012_De_Sanctis_a, 2013_Ammannito_a, 2013_De_Sanctis, 2015_Frigeri_a}. Olivine signatures have been reported in very localized places by \cite{2013_Ammannito_b} and \cite{2014_Ruesch_a}. The nondetection of olivine in the Rheasilvia basin, where it was expected \citep{1997_Binzel, 2017_Raymond} (given the excavation depth that should have been sampled the mantle/lower crust material, as an open window on the mantle composition), adds another constraint on the vertical structure of Vesta's interior and its magmatic history, which is still debated today \citep{2013_McSween_a,2019_McSween}.\par
The presence of OH-bearing dark material, which is associated with a \unit{2.8}{\micro\meter} signature and superimposed over the brighter surface of  Vesta, is inferred to be mainly exogenous and to originate from carbonaceous chondrites \citep{2012_De_Sanctis_b, 2012_McCord, 2012_Reddy_b}. The measurements made by GRaND  also report an important amount of hydrogen in the crust and are consistent with this previous hypothesis; they also confirm that the \ce{Fe}$/$\ce{O} and the \ce{Fe}$/$\ce{Si} ratios of Vesta are compatible with the HED ones, in particular the howardites \citep{2012_Prettyman}.\par
Although the visible wavelengths have been used to study the Vesta mineralogy, instrumental artifacts in the data acquired by the VIR visible channel prevent a complete investigation of the surface of Vesta in the wavelength range (\unit{0.25-1.07}{\micro\meter}). Thanks to the recent development of correction and calibration processes \citep{2020_Rousseau_a, 2020_Rousseau_b}, here we present an analysis of the surface of Vesta in the visible range through different maps of spectral parameters. In Sect. \ref{Sec_Data_Methods} we describe the VIR instrument, the data and associated corrections, the spectral parameters, and the mapping. Resulting maps are described in detail throughout Sect. \ref{Sec_Maps} and are discussed in Sect. \ref{Discussion} together with a study of the lithology based on the VIR visible data.
%
\section{Data and methods}
\label{Sec_Data_Methods}
\subsection{Visible and InfraRed mapping spectrometer}
\label{Subsec_VIR}
The VIR instrument \citep{2011_De_Sanctis_a} is a mapping spectrometer that has been designed to observe Ceres and Vesta surfaces in the visible and infrared wavelengths through two channels. The visible channel ranges from \unit{0.25}{\micro\meter}  to \unit{1.07}{\micro\meter} and the infrared channel between \unit{1.02}{\micro\meter} and \unit{5.09}{\micro\meter}. The instantaneous field of view of VIR (IFOV) is equal to \unit{250}{\micro\radian} $\times$ \unit{250}{\micro\radian} and the spectral sampling in the visible is \unit{1.8}{\nano\meter}/band. Our study is focused on the use of the visible data in the spectral range comprised between $\sim$\unit{380}{\nano\meter} and $\sim$\unit{1000}{\nano\meter}. The different extension of the studied range toward the ultraviolet range compared to that of \cite{2020_Rousseau_a} is made possible thanks to the much higher brightness of Vesta relative to Ceres, which leads to higher signal-to-noise-ratio data in the ultraviolet.\par
\subsection{Data correction}
\label{Subsec_Data_corr}
In this study we use the VIR VISIBLE LEVEL 1B data available on the Planetary Data System (PDS) archive\footnote{\url{https://sbn.psi.edu/pds/resource/dawn/dwncvirL1.html}}. We applied the same level of correction that is described in \cite{2020_Rousseau_a}, to which the reader is referred for more details. Only the photometric correction and the correction of the CCD temperature dependency are different, and these are described in \cite{2021_Scarica} and \cite{2020_Rousseau_b}, respectively. The VIR data with this new level of correction and calibration will be delivered to the PDS and publicly available in the next few months. In this study, we use the calibrated radiance factor (hereafter called reflectance or $I/F$) which is standardized in observation geometry ($\text{incidence}=30\degr$, $\text{emergence}=0\degr$). In addition, in order to decrease the noise level, we applied a median filter with a box width of five spectral channels  to the wavelengths dimension of each spectrum.\par
The data used were acquired over four different mission phases occurring from mid-August 2011 to late July 2012 (see Table \ref{Table1}). A total of 1\,257 hyperspectral cubes are used, corresponding to 16\,122\,723 single observations (or projected pixels) after filtering shadowed areas. Thanks to this dataset, the coverage is almost complete between the south pole and nearly $60\degr$N (see map (e) in Fig. \ref{Fig_Appendix_DENSITY_MAP}, Appendix \ref{Appendix_Density_maps}).
\begin{table}
\caption{Mission phases of Dawn at Vesta used in this study.}
\label{Table1}
\centering
\begin{tabular}{c c c c c}
\hline \hline
\begin{tabular}{@{}l@{}}Mission\\ Phase\end{tabular} & \begin{tabular}{@{}l@{}}Start date\\(mm-dd)\end{tabular} & \begin{tabular}{@{}l@{}}Stop date\\(mm-dd)\end{tabular} & \begin{tabular}{@{}l@{}}Cubes\\(used/total)\end{tabular} & \begin{tabular}{@{}l@{}}Resolution\\ (m/pix)\end{tabular} \\[5pt]
\hline
VSS & 08-12 & 08-29 & 242/271 & 675--716
\\[2.5pt]
VTH & 09-20 & 09-25 & 8/12 & 167--205
\\[2.5pt]
VSH & 09-30 & 10-31 & 325/329 & 165--205
\\[2.5pt]
VH2 & 06-15 & 07-24 & 682/685 & 160--205
\\ \hline
\end{tabular}
\tablefoot{Mission phases are chronologically sorted and we report only the periods during which VIR visible data were acquired. VSS: Vesta Science Survey; VTH: Vesta Transfer to HAMO (High Altitude Mapping Orbit); VSH: Vesta Science High Altitude Mapping Orbit; VH2: Vesta Science High Altitude Mapping Orbit 2. Data were acquired in 2011 for the VSS, VTH and VSH mission phases and in 2012 for the VH2 mission phase. In the fourth column, a discrepancy between processed and available data arises from the occurrence of sky observations or corrupted data. The fifth column provides the approximate minimum and maximum across-track resolutions.}
\end{table}
\begin{figure}
    \includegraphics{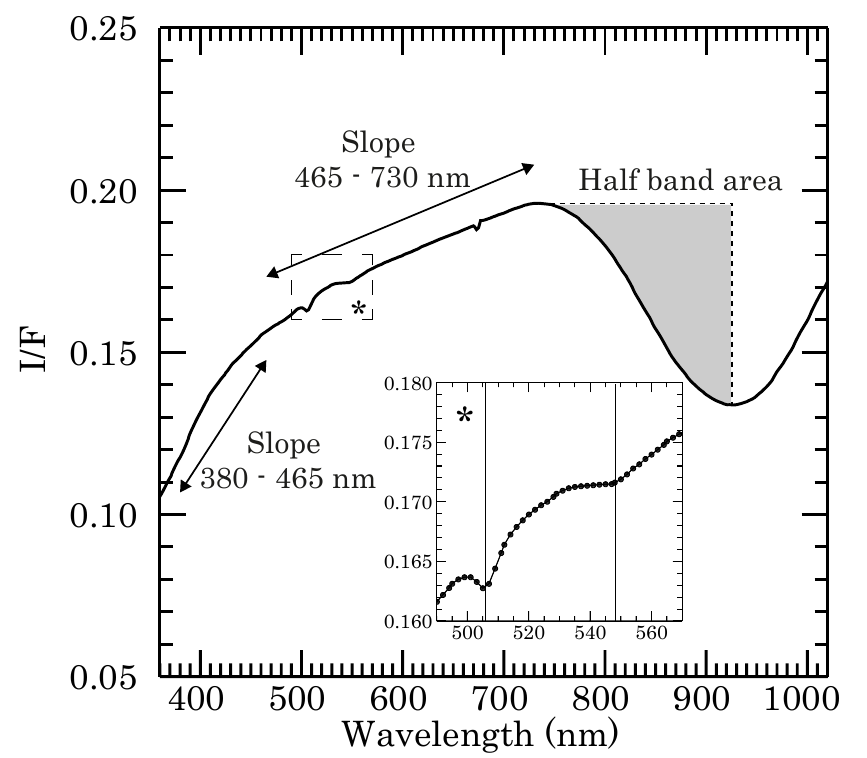}
    \caption{\label{Fig_ref_spectrum}Spectral parameters used in this study reported on the Vesta spectrum. The spectrum is expressed in reflectance at $\text{incidence}=30\degr$ and $\text{emergence}=0\degr$ and is a median of the data acquired during the VH2 orbital phase. The two vertical black arrows on the inset indicate the position of the two $\sim$\unit{506}{\nano\meter} and $\sim$\unit{550}{\nano\meter} spin-forbidden bands. The dots represent the VIR visible spectral channels. The small feature at $\sim$\unit{670}{\nano\meter} is the spectral counterpart of an order-sorting filter.}
\end{figure}
%
\subsection{Vesta VIR visible spectrum}
\label{Subsec_Vesta_spectrum}
The VIR visible spectrum of Vesta is presented in Fig. \ref{Fig_ref_spectrum}. The latter is a median of $8$ million observations acquired during the VH2 mission phase and can be considered as an "average Vesta", given the spatial distribution of the observations during this mission phase (see map (d) in Fig. \ref{Fig_Appendix_DENSITY_MAP}, Appendix \ref{Appendix_Density_maps}).\par
The average spectrum of Vesta is characterized by a clear absorption band centered at about \unit{930}{\nano\meter}. The band center shifts toward shorter or larger wavelengths whether the iron content in the pyroxene is respectively lower or higher \citep{2007_Klima}. A positive slope is observed shortward of the band's left shoulder. This slope has a change at \unit{430-460}{\nano\meter}. The first part, in the far-ultraviolet to near-visible, is characteristic of a \ce{Fe-O} charge transfer absorption \citep{1975_Adams, 1999_Clark}. The second part, which is less positive, ends at \unit{730}{\nano\meter} on the edge of the pyroxene band. Figure \ref{Fig_ref_spectrum} shows a close-up of the \unit{500-560}{\nano\meter} region where two absorption bands at \unit{506}{\nano\meter} and \unit{550}{\nano\meter} can be identified. While the band at \unit{506}{\nano\meter} has previously been identified at Vesta \citep{1997_Golubeva_Shestopalov,1998_Cochran_Vilas,2015_Stephan}, this is the first time that the band at \unit{550}{\nano\meter} can be identified to our knowledge. The bands at \unit{506}{\nano\meter} (SF1) and at \unit{550}{\nano\meter} (SF2) reported in Fig. \ref{Fig_ref_spectrum} are reliable thanks to the recent data correction (see Sect. \ref{Subsec_Data_corr} and \cite{2020_Rousseau_a}) and because of the high number of spectra used to calculate this median spectrum, which allows the removal of the high-frequency noise. It is worth noting that the identification of these two bands on individual spectra can be challenging in low-signal situations. The SF1 band has an intermediate strength, but is narrow and only covers a few VIR spectral channels. Conversely, the SF2 band is wider but very weak, and is therefore often lost in the noise. \cite{2007_Klima} also studied the SF1 and SF2 bands. While the former shifts in wavelength depending on the iron content, they noted that SF2 does not follow the same trend. In Sect. \ref{Subsec_Litho} we address these points further. We did not observe other small absorptions in the VIR spectrum of Vesta. The feature at \unit{670-680}{\nano\meter} is due to an order-sorting filter.
\subsection{Spectral parameters}
\label{Subsec_spectral_parameters}
The spectral variation of Vesta is studied in the \unit{380-1000}{\nano\meter} range. We defined a set of spectral parameters based on the spectrum shown in Fig. \ref{Fig_ref_spectrum} for this purpose.
\subsubsection{Radiance factor and color composites}
The reflectance at \unit{550}{\nano\meter} or $I/F_{550nm}$) is used to characterize the brightness variation of the surface. Three RGB color composites are defined to emphasize the color variations at the surface of Vesta. The first color composite (R1) is based on the red, green, and blue colors attributed to the reflectance at \unit{730}{\nano\meter}, \unit{465}{\nano\meter,} and \unit{380}{\nano\meter}, respectively. The second color composite (R2) is based on the reflectance ratios: \unit{730/550}{\nano\meter} (red), \unit{465/550}{\nano\meter} (green), and \unit{380/550}{\nano\meter} (blue). The third color composite (R3) is similar to the well-known Clementine color composite \citep{1994_Pieters} and is based on the reflectance ratios: \unit{730/465}{\nano\meter} (red), \unit{730/930}{\nano\meter} (green), and \unit{465/730}{\nano\meter} (blue)\footnote{The original Clementine color composite is based on the \unit{750/415}{\nano\meter} (red), \unit{750/950}{\nano\meter} (green), and \unit{415/750}{\nano\meter} (blue) reflectance ratios.}.
\subsubsection{Spectral slopes}
We defined the first spectral slope between \unit{380}{\nano\meter} and \unit{465}{\nano\meter} and the second between \unit{465}{\nano\meter} and \unit{730}{\nano\meter} following equation \ref{eqn:SPECTRAL_SLOPE}. These are expressed in the text as $S_{\lambda_1-\lambda_2}$ (where $\lambda_1$ and $\lambda_2$ are the wavelengths of each side) in units of $\%/\unit{100}{\nano\meter}$.
\begin{equation}
\label{eqn:SPECTRAL_SLOPE}
        S_{\lambda_1-\lambda_2}=
        \frac{(\nicefrac{I}{F})_{\lambda_2}-(\nicefrac{I}{F})_{\lambda_1}}
        {(\nicefrac{I}{F})_{\unit{465}{\nano\meter}}\times(\lambda_2-\lambda_1)}\times10^4
.\end{equation}
\subsubsection{The half band area parameter}
We developed a qualitative index to characterize the BI absorption band. The absorption extent of a band can be estimated with the band area or the band depth. To precisely quantify these indices, the locations of the two wings are generally used to evaluate and to remove the spectral continuum. This step cannot be performed in our study using only the visible data (up to \unit{1000}{\nano\meter}). Therefore, we defined the "half band area" index (HBA) as illustrated in Fig. \ref{Fig_ref_spectrum}. The HBA is based on the estimation of a straight continuum taken between the left wing, defined as the reflectance maximum between \unit{700}{\nano\meter} and \unit{760}{\nano\meter}, and the position of the band minimum\footnote{Here this position does not strictly correspond to the band center because no continuum has been removed; however both quantities should be very close.}, searched for between \unit{900}{\nano\meter} and \unit{950}{\nano\meter}. The classic formula of band area is then applied as follows:
\begin{equation}
\label{eqn:Half_band_area}
       HBA = \int_{\lambda_1}^{\lambda_2} \biggl[1-\frac{(\nicefrac{I}{F})_{\lambda}}{(\nicefrac{I}{F})_{\lambda_1}}\biggr]\, \mathrm{d}\lambda 
,\end{equation}
where $\lambda_1$ and $\lambda_2$ correspond to the wavelength where the smoothed reflectance is the highest in the interval \unit{700-760}{\nano\meter} and the lowest in the interval \unit{900-960}{\nano\meter}, respectively. Because the continuum is defined until the approximate band center, the $HBA$ is not only influenced by the band depth ---linked to the abundance or the grain size--- but also by the band center shift, and then by the pyroxene intrinsic composition \citep{2007_Klima}. Consequently, variations of the half band area only give a qualitative hint that BI is changing. This index allows a comparison to be made with caution with other studies dealing with infrared data.
\subsubsection{Band centers}
Monitoring of the band center allows the mineral chemistry variation to be studied. At Vesta, a pyroxene lithology can be associated with the observed terrains (see e.g. \cite{2013_McSween_a,2013_Ammannito_a,2015_Stephan}) and usually the band centers of the pyroxene bands are used. In Sect. \ref{Discussion}, we address the lithology of Vesta resulting from the study of the BI, SF1, and SF2 band centers. In this analysis, the position of the BI band minimum is adopted as the band center. Regarding the SF1 and SF2 bands, their wings are fixed for SF1 at \unit{501}{\nano\meter} and \unit{512}{\nano\meter} and for SF2 at \unit{535}{\nano\meter} and \unit{560}{\nano\meter}). Their band centers are calculated using a Gaussian function to fit the continuum-removed spectra. Further details are provided in Appendix \ref{Appendix_BC_method} and are illustrated in Fig. \ref{Fig_Appendix_SF_band_fit}. The maps of the band centers are also provided in Appendix \ref{Appendix_maps_band_center} and Figs. \ref{Fig_Appendix_BI_map}, \ref{Fig_Appendix_SF1_map}, and \ref{Fig_Appendix_SF2_map}. Contrary to BI and SF1, the spatial distribution of the SF2 band center is meaningless and consequently unsuitable for scientific purposes, as detailed in Sect. \ref{Subsec_Litho}.
%
\subsection{Map projections}
\label{Subsec_maps_proj}
The methodology used to build the maps presented in Sect. \ref{Sec_Maps} is the same as introduced in \cite{2020_Rousseau_a}, to which the reader is referred for more details. The same Mollweide projection is used and we add a mask above $60\degr$N because of the partial to total lack of data above this latitude.\par
In the following section, we present the $I/F_{550nm}$ and RGB maps (Figs. \ref{Fig_IoF_550}, \ref{Fig_RGB_380_465_730}, \ref{Fig_Ratio_550} and \ref{Fig_Clementine}), while the spectral slopes and band area maps (Fig. \ref{Fig_Slope_380_465}, \ref{Fig_Slope_465_730} and \ref{Fig_Half_BA}) are superimposed with an opacity of $70\%$ on a Framing Camera mosaic \citep{2015_Roatsch}. The map of the Framing Camera is presented in Appendix \ref{Appendix_HAMO} and the maps of the spectral slopes and band area are presented without FC background in Appendix \ref{Appendix_maps_no_transp} (see Figs. \ref{Fig_Appendix_SLOPE_380_465}, \ref{Fig_Appendix_SLOPE_465_730} and \ref{Fig_Appendix_Half_BA}).\par
Contrary to the north pole, the south pole of Vesta was illuminated, and thus imaged, during the Dawn mission. In Appendix \ref{Appendix_South_pole}, we present (see Fig. \ref{Fig_Appendix_South_pole}) a panel of various stereographic views centered on the south pole and extending to $\sim55\degr$S, through the same spectral indicators introduced in Sect. \ref{Subsec_spectral_parameters}.\par
It is worth noting that some artifacts are still present on the maps: (a) The VIR maps lack some data; these areas appear white, in particular around the latitude $\sim50\degr$N and also for example  at $17\degr$N-$13\degr$E, $10\degr$N-$156\degr$E as well as close to the south pole. (b) There are also artifacts due to a combination of a lack of data and a less efficient photometric correction in the VSS mission phase, which are visible around the latitude $30\pm10\degr$N and particularly at the longitudes $60\degr$E, $100\degr$E, $210\degr$E, $240\degr$E and $315\degr$E. (c) A less efficient photometric correction for the longest wavelengths causes some cube footprints to be visible on the maps that use the channel around \unit{930}{\nano\meter}, that is, the Clementine composite and half-band area maps in Figs. \ref{Fig_Clementine}, \ref{Fig_Half_BA}, and \ref{Fig_Appendix_Half_BA}. These adverse effects must be taken into account for a clear interpretation of the maps.\par
Features of Vesta discussed hereafter are referred to by their names and reference numbers (in parenthesis) in the text, while being identified only by their numbers on the maps. Names, numbers, and coordinates are reported in Table \ref{Table2}.
\begin{table}
\caption{Main features mentioned in the text.}
\label{Table2}
\centering
\begin{tabular}{c c c c}
\hline \hline
\begin{tabular}{@{}l@{}}\#\end{tabular} &
\begin{tabular}{@{}l@{}}Vesta surface\\formation names\end{tabular} & \begin{tabular}{@{}l@{}}Longitude\end{tabular} & \begin{tabular}{@{}l@{}}Latitude\end{tabular}\\[5pt]
\hline
1 & Rubria & $18\degr$E & $7\degr$S
\\[2.5pt]
2 & Occia & $18\degr$E & $15\degr$N
\\[2.5pt]
3 & Tarpeia & $29\degr$E & $70\degr$S
\\[2.5pt]
4 & Bellicia & $48\degr$E & $38\degr$N
\\[2.5pt]
5 & Arruntia & $72\degr$E & $39\degr$N
\\[2.5pt]
6 & Matronalia Rupes\tablefootmark{a} & $83\degr$E & $49\degr$S
\\[2.5pt]
7 & Lollia & $92\degr$E & $37\degr$S
\\[2.5pt]
8 & Serena & $121\degr$E & $20\degr$S
\\[2.5pt]
9 & Severina & $122\degr$E & $75\degr$S
\\[2.5pt]
10 & Octavia & $147\degr$E & $3\degr$S
\\[2.5pt]
11 & Aricia Tholus\tablefootmark{b} & $62\degr$E & $13\degr$N
\\[2.5pt]
12 & Marcia & $190\degr$E & $9\degr$N
\\[2.5pt]
13 & Tuccia & $197\degr$E & $40\degr$S
\\[2.5pt]
14 & Antonia & $201\degr$E & $59\degr$S
\\[2.5pt]
15 & Cornelia & $226\degr$E & $9\degr$S
\\[2.5pt]
16 & Vibidia & $220\degr$E & $27\degr$S
\\[2.5pt]
17 & Fabia & $266\degr$E & $16\degr$N
\\[2.5pt]
18 & Teia & $271\degr$E & $3\degr$S
\\[2.5pt]
19 & Canuleia & $295\degr$E & $34\degr$S
\\[2.5pt]
20 & Charito & $301\degr$E & $45\degr$S
\\[2.5pt]
21 & Oppia & $309\degr$E & $8\degr$S
\\[2.5pt]
22 & Justina & $318\degr$E & $34\degr$S
\\[2.5pt]
\\ \hline
\end{tabular}
\tablefoot{First column: identification number reported on the maps of Sect. \ref{Sec_Maps}; these are ordered from west to east and from north to south. Second column: formation name. Third and fourth columns: longitude and latitude. Most of the formation are impact craters with a few exceptions: \tablefoottext{a}{Matronalia Rupes is a long scarp defining the eastern rim of the Rheasilvia basin \citep{2012_Jaumann,2014_Krohn}.}  \tablefoottext{b}{Aricia Tholus is a dome and could be a remnant of an impact basin rim or of volcanic origin (dike-like) \citep{2014_Williams_b}.}
}
\end{table}
%
%
%
\section{Global maps of spectral parameters}
\label{Sec_Maps}
\subsection{Map of reflectance at \unit{550}{\nano\meter}}
\label{SubSec_IoF_550}
\begin{figure*}
    \centering
    \includegraphics[width=17cm]{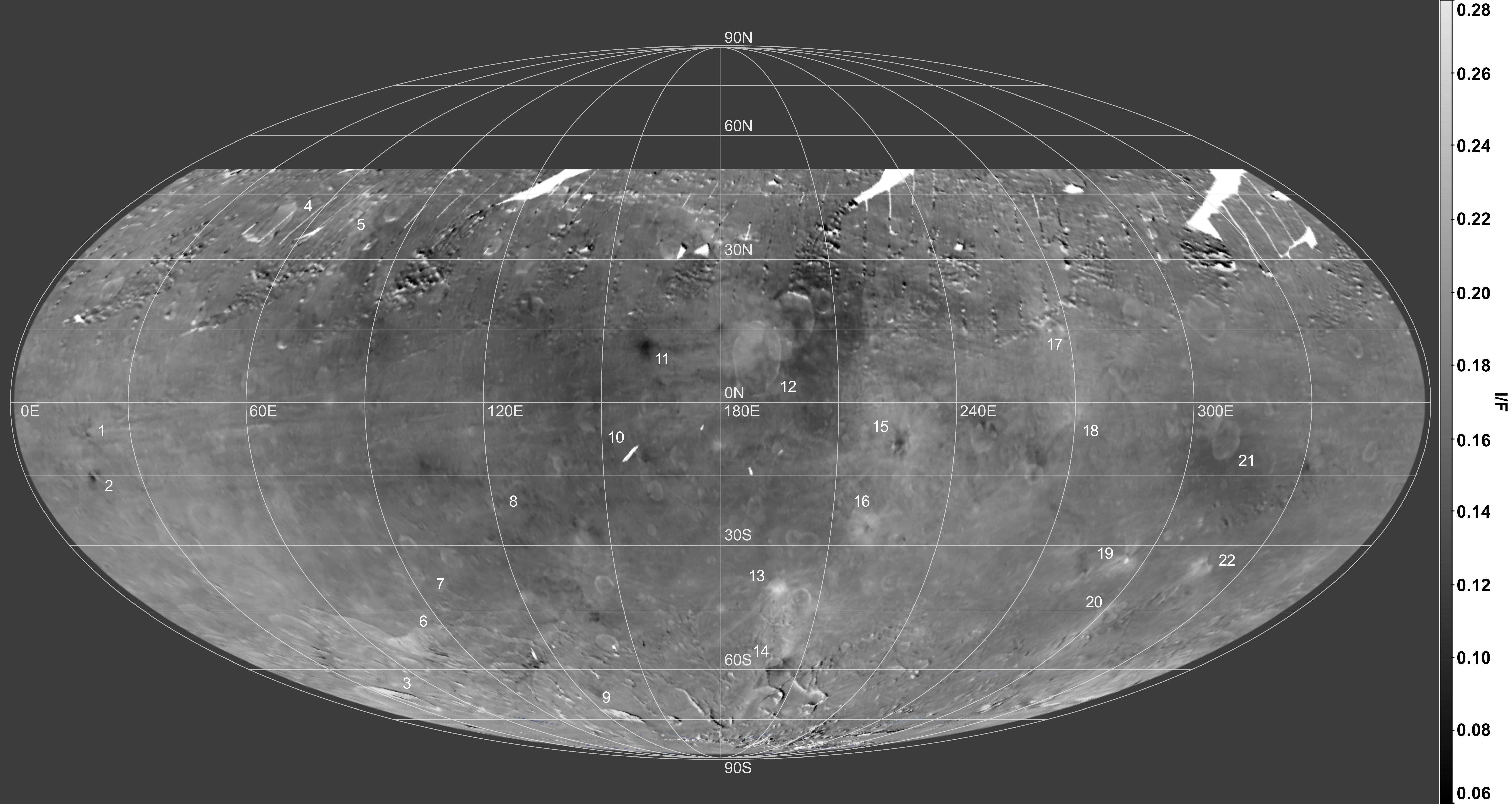}
    \caption{\label{Fig_IoF_550} Map of the VIR reflectance at \unit{550}{\nano\meter}. Here and in Figs. \ref{Fig_RGB_380_465_730} to \ref{Fig_Half_BA}, and \ref{FIG_Lithology_1_map}, numbers refer to the features of Table 2, and white areas correspond to missing data or overexposed spots.}
\end{figure*}

The VIR map of the reflectance at \unit{550}{\nano\meter} is presented in Fig. \ref{Fig_IoF_550}. To highlight its variation at the surface of Vesta, the reflectance range has been fixed between $0.06$ and $0.28$, which includes more than $98\%$ of the dataset. The surface of Vesta is relatively bright, with a mean $I/F_{550nm}$ of $0.175\pm0.026$\footnote{All the errors are given as the standard deviation of the distribution in the $0.06-0.28$ range} and, as mentioned by \cite{2012_Reddy_a}, has the largest variation of albedo among the known asteroids, including the recently visited Bennu and Ryugu \citep{2020_DellaGiustina,2019_Watanabe,2019_Sugita}. Reflectance maps and the main features on  Vesta have  already been thoroughly studied based on the high-spatial-resolution Framing Camera data; first by \cite{2012_Reddy_a} who used data acquired during the approach phase, that is, with a moderate resolution, and then by \cite{2013_Li_a} and \cite{2013_Schroder} who used data of higher resolution and detailed the global and resolved photometric properties of the surface, respectively. These studies all used data at or close to \unit{750}{\nano\meter} but our VIR reflectance map, using data at \unit{550}{\nano\meter}, shares globally the same characteristics with surface variation visible at different scales. In particular, a hemispherical dichotomy is observable with a region comprised between $60\degr$E and $220\degr$E which is darker ($\overline{I/F}_{550nm}=0.163\pm0.023$) than the other part of the surface ($\overline{I/F}_{550nm}=0.179\pm0.022$), excluding the latitude below $\sim50\degr$S which appears to be brighter too ($\overline{I/F}_{550nm}=0.194\pm0.028$). Local exceptions are noticeable, as in Oppia crater (21) and its surroundings, or the small Rubria (1) and Occia (2) craters. Cornelia crater (15), characterized by close dark and extended bright ejecta, marks the eastern boundary of the dichotomy. Among the darkest longitudes, the north of Marcia crater (12) and Tuccia crater (13) appear to be brighter than the close terrains. Tuccia (13) is also the brightest area on Vesta's surface ($\overline{I/F}_{550nm}=0.242\pm0.024$; in Fig. \ref{Fig_IoF_550}) while the darkest are Occia (2) and the unnamed crater on Aricia Tholus (11) (with $\overline{I/F}_{550nm}$ around $0.131\pm0.030$ and $0.107\pm0.017$, respectively).
%
%
\subsection{Color composite maps}
\label{SubSec_RGB_380_465_730}
\begin{figure*}
    \centering
    \includegraphics[width=17cm]{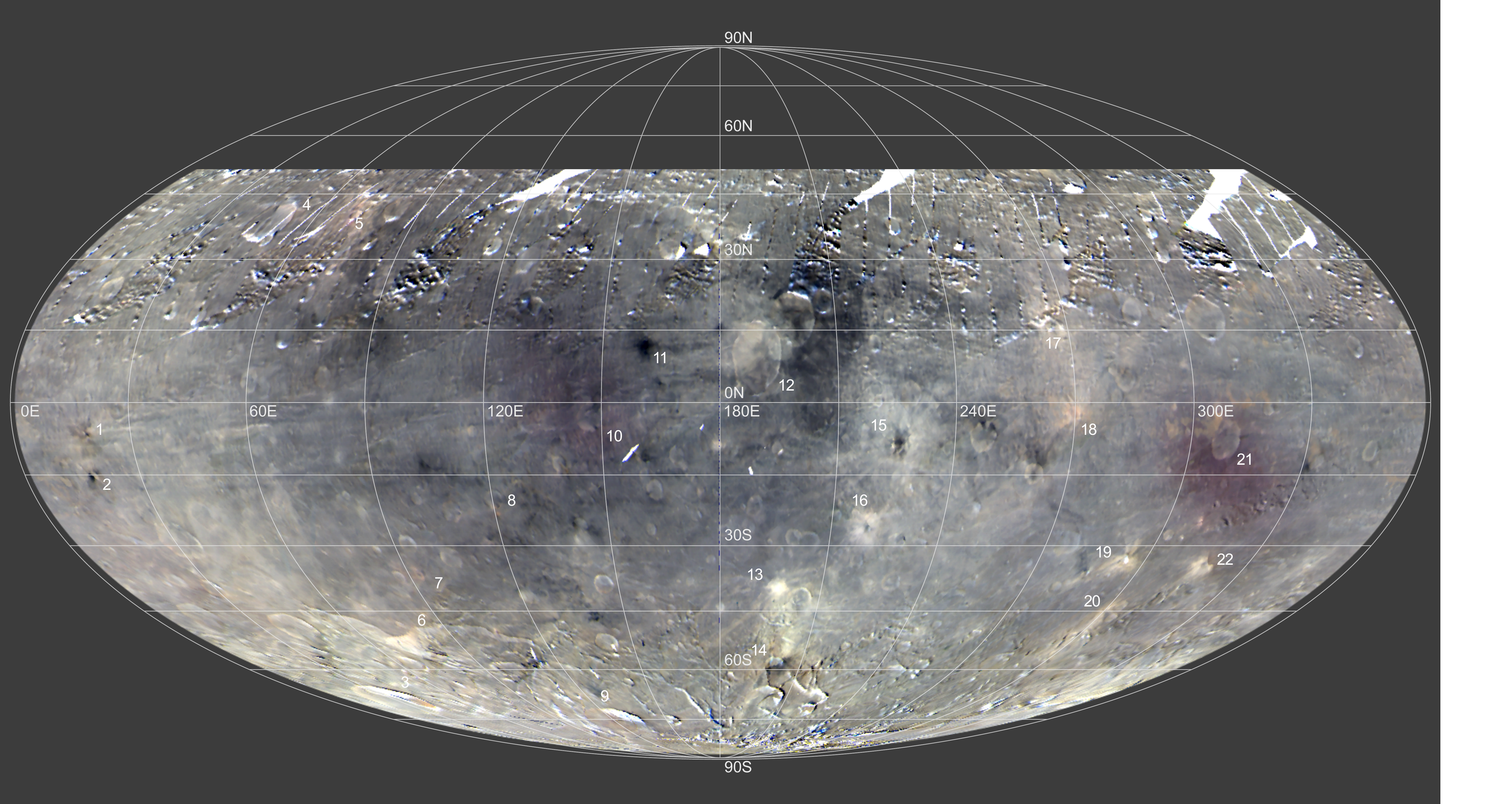}
    \caption{\label{Fig_RGB_380_465_730} VIR R1 color composite map using the reflectance at \unit{730}{\nano\meter}, \unit{465}{\nano\meter,} and \unit{380}{\nano\meter} for the red, green, and blue channels, respectively.}
\end{figure*}

\begin{figure*}
    \centering
    \includegraphics[width=17cm]{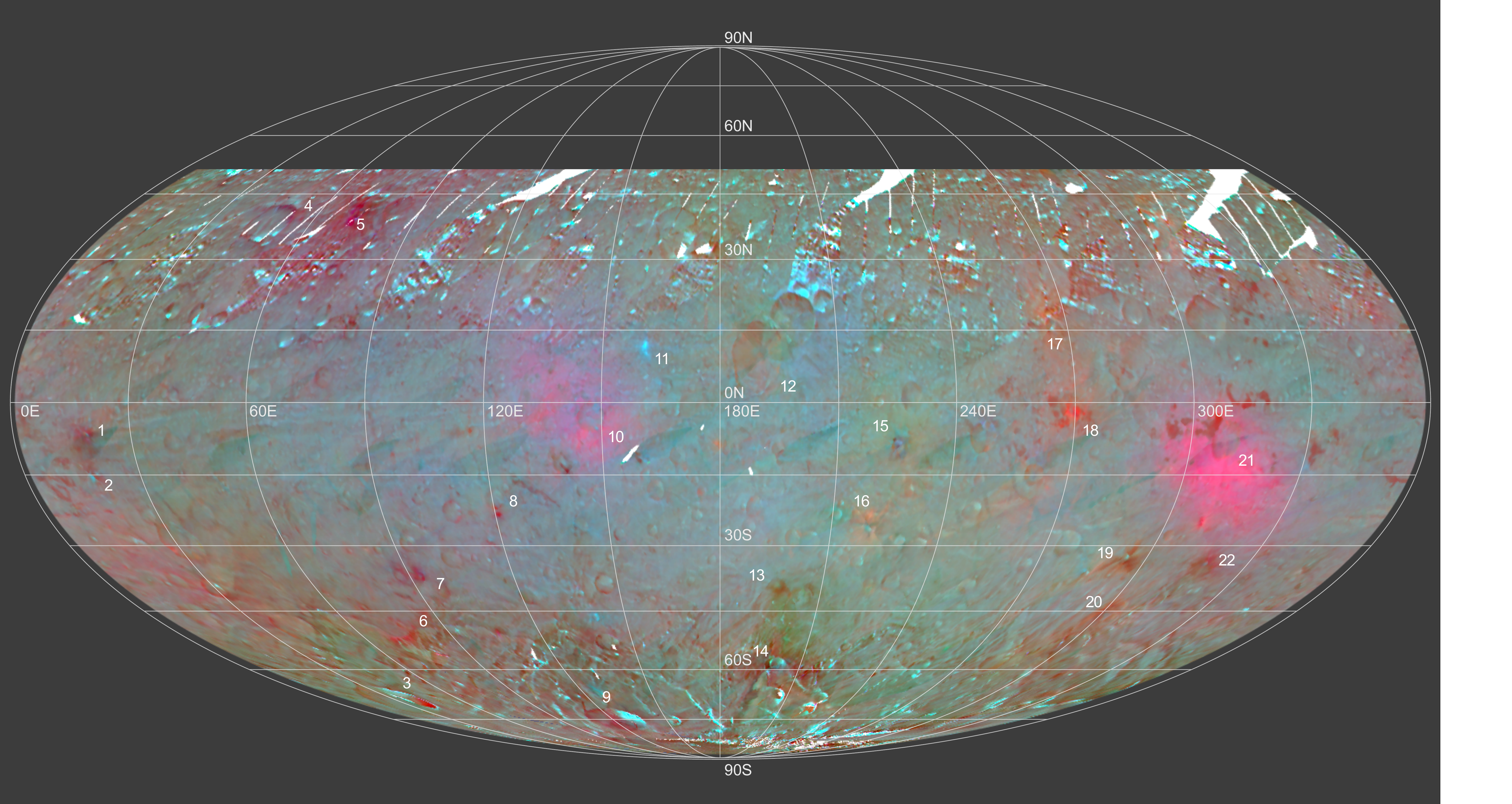}
    \caption{\label{Fig_Ratio_550} VIR R2 color composite map using the reflectance ratios \unit{730/550}{\nano\meter}, \unit{465/550}{\nano\meter,} and \unit{380/550}{\nano\meter} for the red, green, and blue channels, respectively.}
\end{figure*}

\begin{figure*}
    \centering
    \includegraphics[width=17cm]{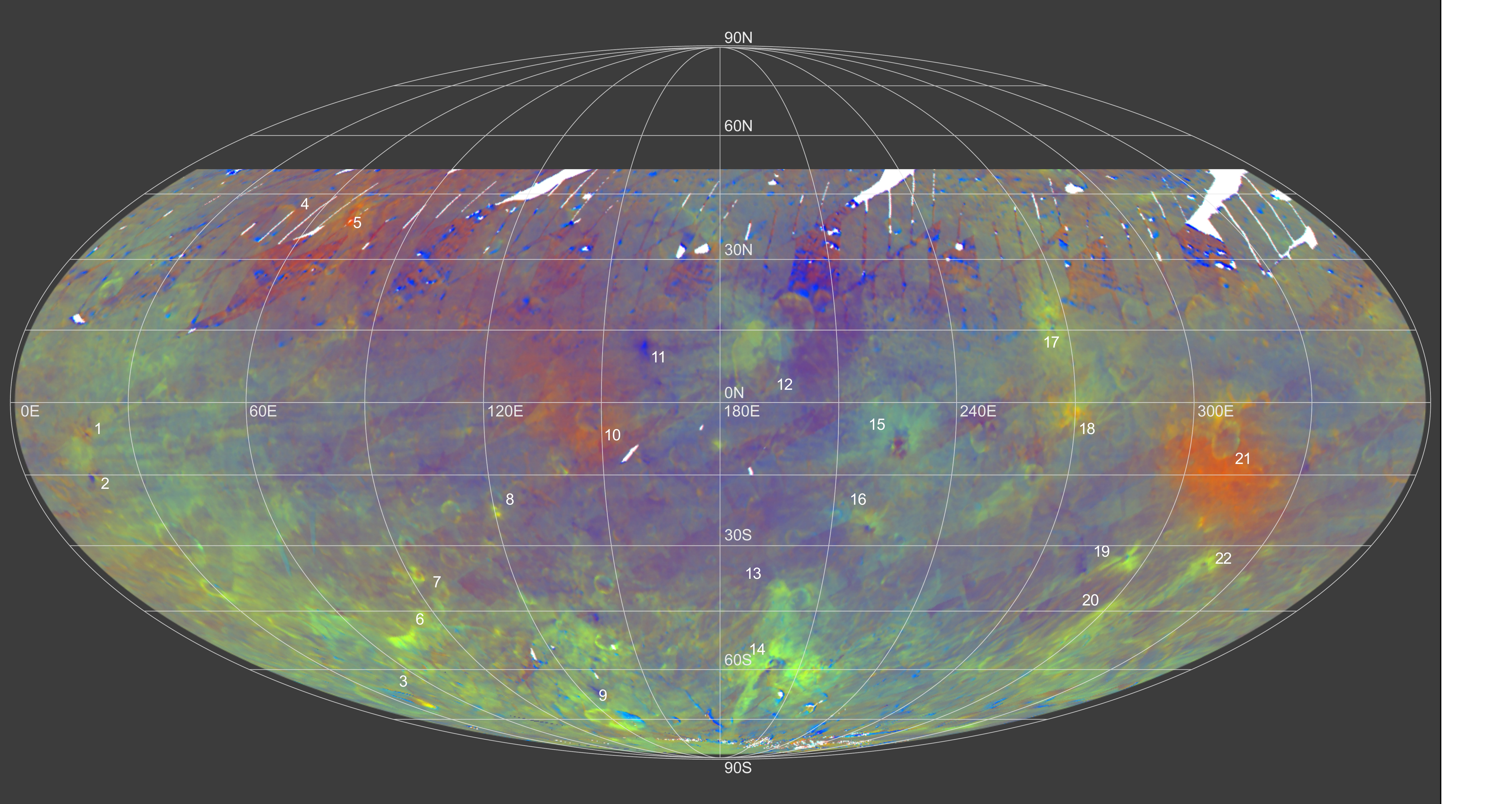}
    \caption{\label{Fig_Clementine} VIR R3 color composite map using the reflectance ratio at \unit{730/465}{\nano\meter}, \unit{730/930}{\nano\meter,} and \unit{465/730}{\nano\meter} for the red, green, and blue channels, respectively (similar to Clementine color ratios).}
\end{figure*}

We present diverse maps based on color composites through Figs. \ref{Fig_RGB_380_465_730}, \ref{Fig_Ratio_550}, and \ref{Fig_Clementine}. Figure \ref{Fig_RGB_380_465_730} presents the R1 composite which combines the information of both reflectance and color variations. The R2 composite is shown in Fig. \ref{Fig_Ratio_550} and is based on the same reflectance values as R1 but normalized by ${I/F}_{550nm}$ to get rid of reflectance variations. Finally,  representing R3, Fig. \ref{Fig_Clementine} uses the well-known Clementine color ratio composite devised by \cite{1994_Pieters}. The map highlights in red the visible slope that is larger than the one represented in blue and shows the significance of the BI  in green \citep{2012_Reddy_a}. The R3 map is affected by artifacts; these can be seen as discontinuities along the seams where the cube footprints are merged, which are likely caused by a less efficient photometric correction towards the longest wavelengths.\par 
At a large scale, the green color in the R3 composite (Fig. \ref{Fig_Clementine}) predominates in the south pole (see also Fig. \ref{Fig_Appendix_South_pole}) below $45\degr$S. This can be explained by the Rheasilvia and Veneneia basins (see their limits in Fig. \ref{Fig_Appendix_HAMO_MAP}) and the associated geological structure; for example Matronalia Rupes (6). A long, diffuse, and greenish latitudinal band is also visible between $\sim0\degr$E and $\sim60\degr$E, with a slight east--west orientation from the south, and could correspond to the Rheasilvia ejecta \citep{2012_Reddy_a,2013_Ammannito_a}. This band is barely seen on the R1 and R2 composite maps despite the correlation between green and yellowish terrains in the R3 and R1 composites, respectively. The dichotomy observed in the ${I/F}_{550nm}$ map is visible through the R2 and R3 composites. The area between $60\degr$E and $220\degr$E is bluer for the former and more violet for the latter. This characteristic is visible on the R1 map but without the possibility to disentangle the reflectance from the color variations.\par
The R1 composite (Fig. \ref{Fig_RGB_380_465_730}) reveals that the reflectance and the color variations are not always correlated at local scale. However, as observed for the reflectance, color changes are generally associated with impact craters. The two unique large and violet areas associated with Octavia (10) and Oppia (21) craters illustrate this correlation. These features are clearly identifiable with the R2 and R3 composites as well (Figs. \ref{Fig_Ratio_550} and \ref{Fig_Clementine}), where they appear pink and orange, respectively. The R1 composite also highlights Fabia (17), Teia (18), Canuleia (19), Charito (20), and Justina (22) craters (and to a lesser extent Marcia (12) and Vibidia (16)), all showing a light brown color. These craters have a color similar to Arruntia (5) crater, whereas they do not share the same color in the R2 and R3 composites, where they present orange-red and green tones, respectively. Based on the significance of the R3 composite bands, a green color is indicative of a deeper BI. The northern rims of Marcia crater (12) (which takes place within the aforementioned globally bluer area) is well visible in green through the R3 composite. Surprisingly, the R2 composite highlights all of these differences (as Arruntia (5) differs in color from those craters) while not based on the depth of the BI. Occia (2) ejecta and the crater on Aricia Tholus (11) (i.e., darkest features on Vesta) are distinctly blue ---even the bluest--- in the R2 and R3 composites. On the R3 map, we also see that Arruntia (5) crater, and to a lesser extent Rubria (1) crater, appear orange as well. This denotes a relatively steep slope in the visible, while they are not sharing the same color as Occia (2) and Octavia (10) craters in R1 and R2, where Arruntia (5) and Rubria (1) craters appear rather light brown and reddish, respectively. Finally, Cornelia crater (15) shows a peculiar behavior; through the R2, its proximal ejecta (dark in R1) appear blue to violet and the distal ejecta (bright in R1) green to orange. The contrast is larger through the R3, with the former ejecta showing a blue to purple color and the latter a light green tone. 
%
%
\subsection{Map of the $S_{380-465nm}$ slope}
\label{SubSec_Slope_380_465}
\begin{figure*}
    \centering
    \includegraphics[width=17cm]{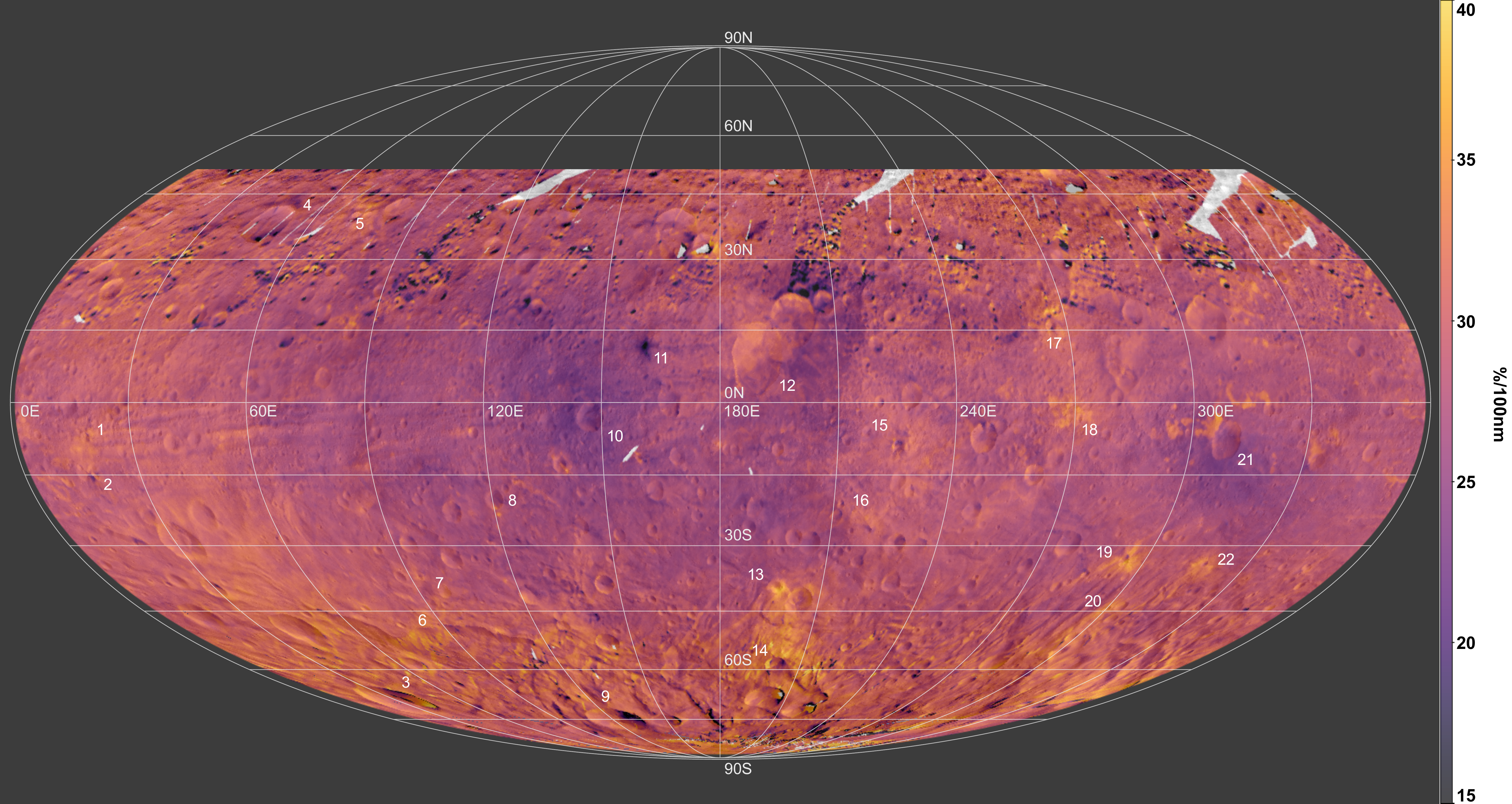}
    \caption{\label{Fig_Slope_380_465} Map of the $S_{380-465nm}$ spectral slope (\unit{380-465}{\nano\meter}) superimposed on the FC HAMO map (see Sect. \ref{Subsec_maps_proj}).}
\end{figure*}

Figure \ref{Fig_Slope_380_465} shows the variation of the $S_{380-465nm}$ value at the surface of Vesta (the reader is also referred to Fig. \ref{Fig_Appendix_SLOPE_380_465} for a version of this map without any layout). The color scale encompasses the whole significant range of the $S_{380-465nm}$ values between $15\%/\unit{100}{\nano\meter}$ and $40\%/\unit{100}{\nano\meter}$ (more than $98\%$ of the data), although only a few areas show effectively extreme values of $S_{380-465nm}$.\par
On the global scale we again encounter the hemispherical dichotomy as observed on the reflectance and color composite maps. However, the contrast is fainter towards the eastern limit (around $220\degr$E). Towards the west (around $60\degr$E), the limit is shown to some extent by what could correspond to the Rheasilvia ejecta. The south pole of Vesta, as visible in Fig. \ref{Fig_Appendix_South_pole}, shows higher $S_{380-465nm}$ than the rest of the surface on average, with a mean $S_{380-465nm}$ of around $31.3\pm3.2\%/\unit{100}{\nano\meter,}$ against $28.6\pm2.8\%/\unit{100}{\nano\meter}$ at higher latitudes. This emphasizes again the influence of the Rheasilvia and Veneneia basins.\par
At local scale, the highest and the lowest $S_{380-465nm}$ values are observed for Tuccia crater (13) and for the crater located on Aricia Tholus (11), respectively. Interestingly, these are two fresh craters that have probably excavated or deposited different types of material on the surface. Other craters showing high $S_{380-465nm}$ are visible on Vesta, as previously noticed on the composite maps (e.g., Fabia (17), Teia (18), Canuleia (19), Charito (20), and Justina (22) craters). Another feature, associated with the northwestern ejecta blanket of Oppia crater (21) in all likelihood, contrasts with the surrounding ejecta and indicates certainly an important change in the properties of the terrains \citep{2015_Tosi}. Conversely, at a regional scale, the southern ejecta of Oppia crater (21) show low $S_{380-465nm}$ values. The same low $S_{380-465nm}$ is observed for Octavia (10) crater, thus recalling the analysis made with the color composite maps where those two craters show similar colors.
%
%
\subsection{Map of the $S_{465-730nm}$ slope}
\label{SubSec_Slope_465_730}
\begin{figure*}
    \centering
    \includegraphics[width=17cm]{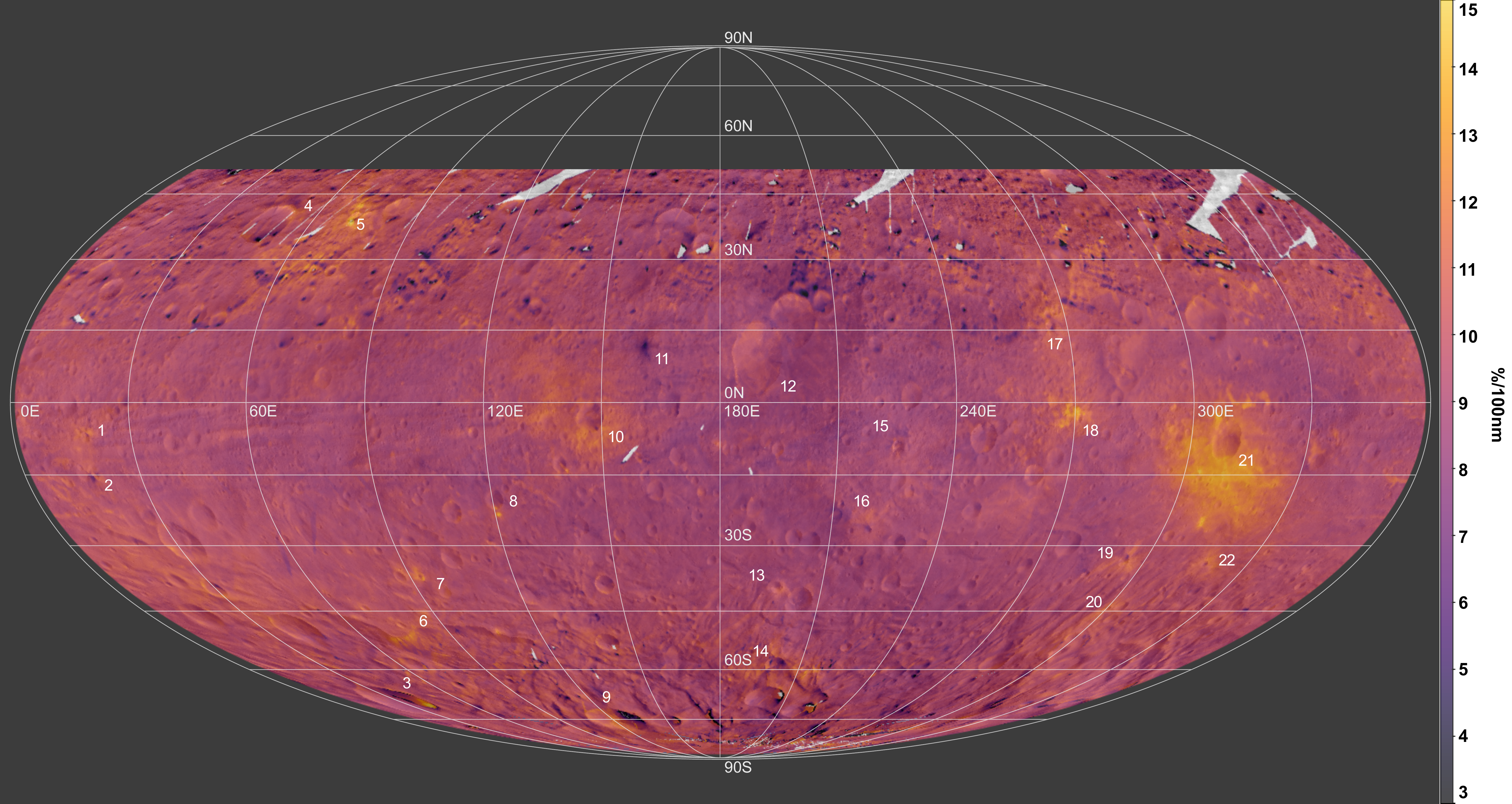}
    \caption{\label{Fig_Slope_465_730} Map of the $S_{465-730nm}$ spectral slope (\unit{465-730}{\nano\meter}) superimposed on the FC HAMO map (see Sect. \ref{Subsec_maps_proj}).}
\end{figure*}

The spectral slope at intermediate VIR visible wavelengths ($S_{465-730nm}$) presents a globally more homogeneous distribution compared to $S_{380-465nm}$, as visible in Figs. \ref{Fig_Slope_465_730} and \ref{Fig_Appendix_SLOPE_465_730}. The surface of Vesta has a mean $S_{465-730nm}$ of $9.5\pm1.0\%/\unit{100}{\nano\meter}$. As opposed to the other indicators, no obvious variations are noticeable at large scale, and only four to five features clearly stand out. This is the case for  Oppia (21) and Octavia (10) craters, which show an analogous trend through the various spectral indicators. Similarly, Teia (18) and Arruntia (5) craters exhibit high $S_{465-730nm}$ values. This latter and the southern ejecta of Oppia (21) are the areas on the surface with the highest value of $S_{465-730nm}$ (around $12.0\pm1.0\%/\unit{100}{\nano\meter}$). Conversely, the crater on top of Aricia Tholus (11) appears to be the feature with the lowest $S_{465-730nm}$ value ($6.5\pm1.1\%/\unit{100}{\nano\meter}$. Other small craters are slightly recognizable on Vesta through the $S_{465-730nm}$ map (\ref{Fig_Slope_465_730}). 
%
%
\subsection{Map of the band area}
\label{SubSec_Half_BA}
\begin{figure*}
    \centering
    \includegraphics[width=17cm]{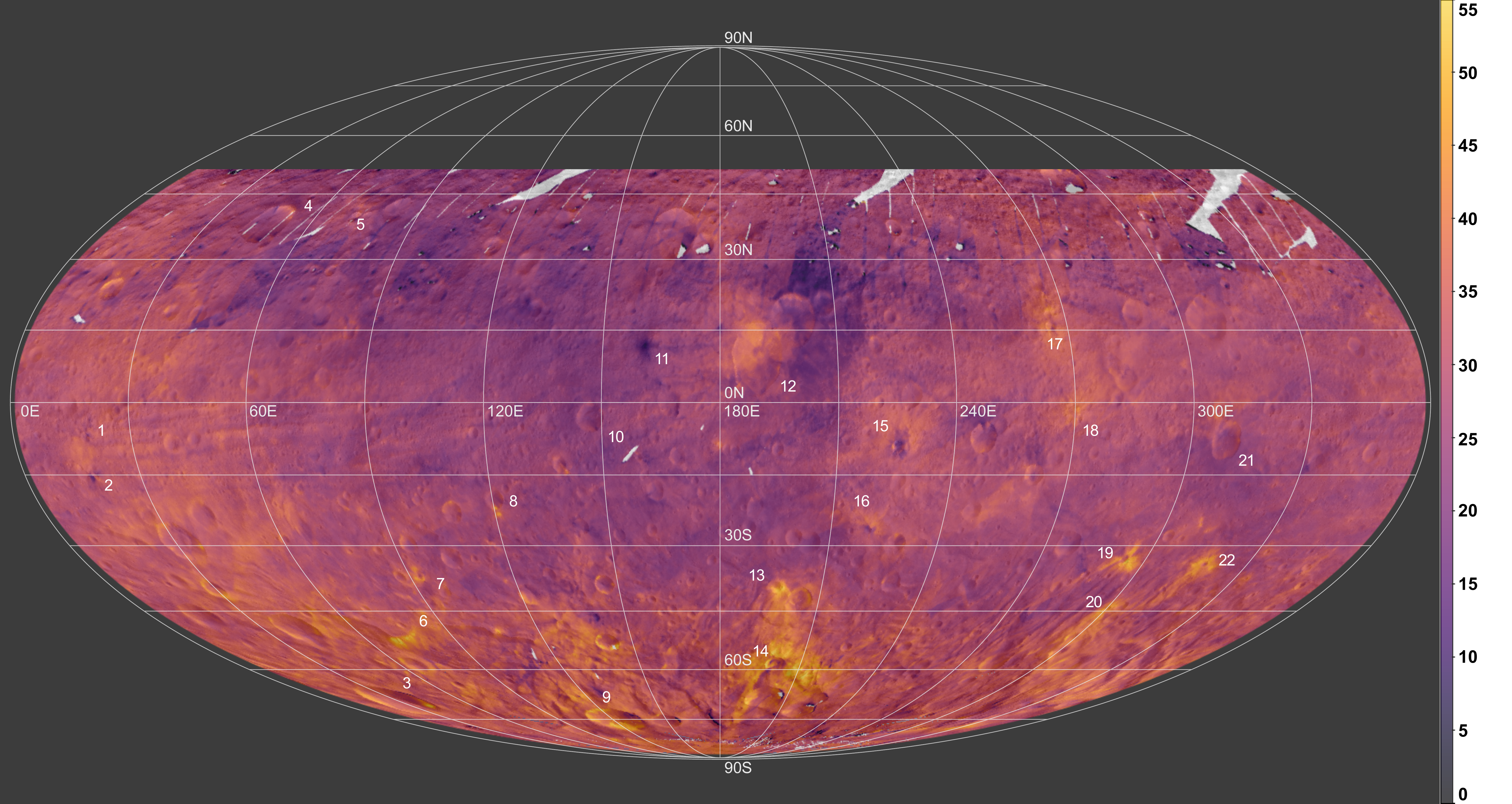}
    \caption{\label{Fig_Half_BA} Map of the half band area around \unit{930}{\nano\meter} superimposed on the FC HAMO map (see Sect. \ref{Subsec_maps_proj}).}
\end{figure*}

The half band area (HBA) spectral parameter, as defined in Sect. \ref{Subsec_spectral_parameters}, helps to qualitatively characterize the pyroxene absorption band. As shown in the maps from \ref{Fig_Half_BA} and \ref{Fig_Appendix_Half_BA}, this parameter displays some artifacts, similar to the R3 composite but weaker. Beyond this, well-contrasted variations can be observed at various scales. The region between $60\degr$E and $220\degr$E also highlighted in the reflectance, color composites, and $S_{380-465nm}$ maps (Figs. \ref{Fig_IoF_550} to \ref{Fig_Slope_380_465}, respectively), appears to have a low HBA. We also observe that the south pole (Fig. \ref{Fig_Appendix_South_pole}) has a high HBA as well as the longitudes from $\sim0\degr$E to $\sim60\degr$E. Tuccia (13) and Antonia (14) craters and ejecta show a very high HBA, probably the highest on Vesta's surface. Other features stand out, such as Canuleia (19), Charito (20), and Justina (22) craters. This is also the case for the distal ejecta of Cornelia (15). A part of Matronalia Rupes (6)  also presents a high HBA. To a lesser extent, the northern inner ground of Marcia crater (12) is visible as well as Fabia crater (17). As for the other spectral parameters, the crater on top of Aricia Tholus (11) is highly visible, showing a low HBA. While presenting a peculiar analogous behavior on the previous maps, Octavia (10) and Oppia (21) craters do not stand out from their surroundings in the HBA map.
%
\section{Discussion}
\label{Discussion}
The composition of Vesta has been thoroughly investigated based on various studies using the VIR data and using different methods. The excellent imaging capability of  VIR combined with a broad spectral range allow observation of the geological context and thus mineralogical investigation: spectra and derived spectral parameters were mapped on Vesta with an unprecedented spatial resolution. The derived lithological maps ---based on the variation of band centers--- are described by \cite{2013_Ammannito_a}, while comparisons of spectral data and the different classes of HED meteorites have been performed by \cite{2013_De_Sanctis}. In Sect. \ref{Subsec_Litho}, we discuss the derivation of the lithology on the basis of the VIR visible data.\par 
The interpretation of spectral slopes and color composites is not as direct as for other parameters (e.g., band area, band center). Grain size variations, composition, mixing modalities, porosity, and space weathering effects may all cause changes of spectral slopes (e.g., \cite{1974_Adams,2013_Cloutis,2016_Pieters,2016_Poch_a,2020_Sultana}). In addition, interpretation of the HBA parameter introduced in this study must be done carefully. This parameter is modified by band center shifts (influenced by the mineral composition \citep{1974_Adams}) and band depth variations (influenced by grain size and mineral abundances \citep{1970_Hunt_Salisbury,2013_Cloutis}). Nonetheless, the diversity of the proxies used here allows us to emphasize similarities and differences among features of interest in light of previous studies. In Sect. \ref{SubSec_impact's_impacts} we highlight the role of the impact craters on the spectral diversity observed at Vesta. Finally, we address the olivine-rich spots identified in previous studies in Sect. \ref{SubSec_olivine}.\par
\subsection{Lithology}
\label{Subsec_Litho}
The pyroxene band centers tend to shift depending on the composition of the mineral. In the case of pyroxene, the BI and BII centers are linked to the \ce{Ca} and \ce{Fe} content \citep{1972_Adams,1974_Adams,1987_Aoyama,1991_Cloutis,1993_Burns,2007_Klima}. To further study the composition of Vesta, one can combine spectral indices derived from HED meteorites and VIR data. Therefore, computing the band center allows us to characterize the surface lithology. \cite{2012_De_Sanctis_a}, \cite{2013_McSween_a}, \cite{2013_Ammannito_a}, and \cite{2015_Stephan} performed this investigation by taking advantage of the BI and BII centers, and hence by partially relying on the VIR infrared data. Here we discuss the extent to which the same method can be applied using visible data alone.\par
\begin{figure}
    \includegraphics{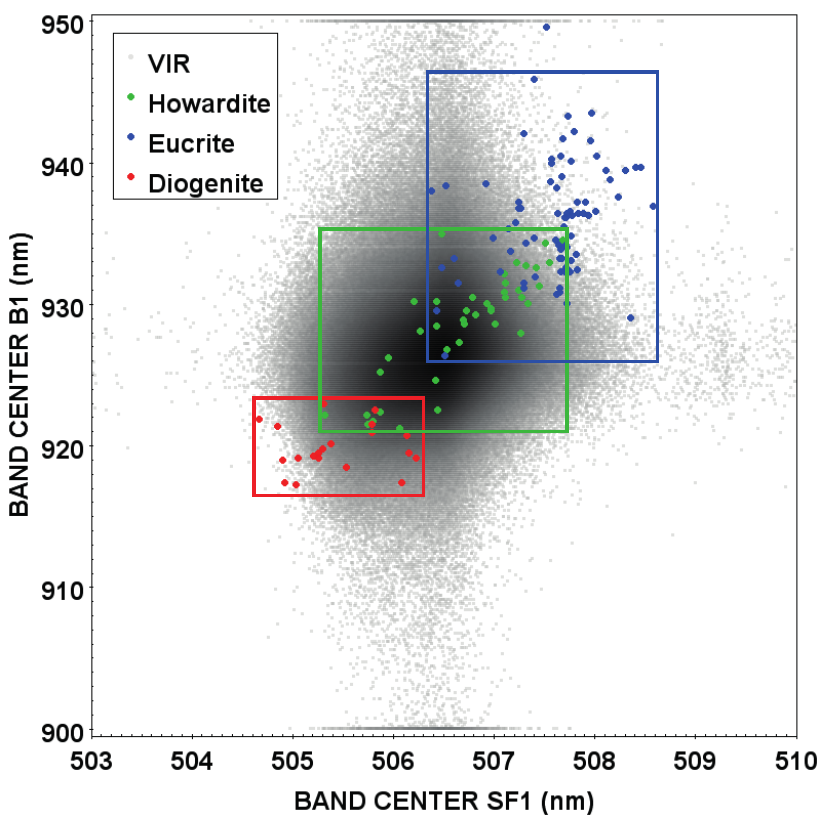}
    \caption{\label{Fig_Scatterplot_SF1_BI} SF1 center versus BI center positions. The gray-black cloud corresponds to the VIR data and the colored dots to HED data. The boxes define the lithology attributed to the VIR data based on the band centers of the HED.}
\end{figure}
\begin{figure*}
    \centering
    \includegraphics[width=17cm]{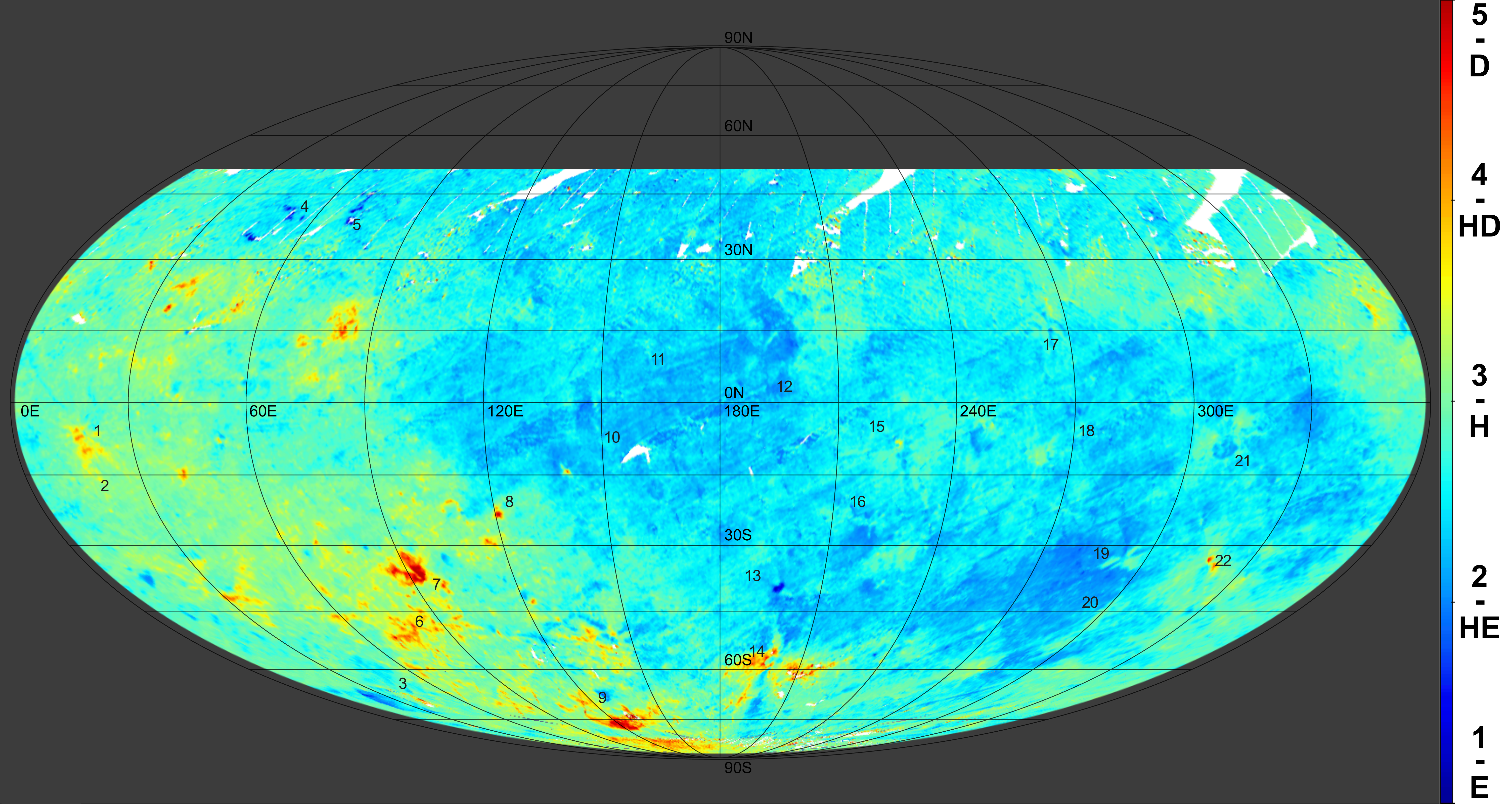}
    \caption{\label{FIG_Lithology_1_map}Map of the lithology based on the bands at \unit{505}{\nano\meter} and \unit{930}{\nano\meter}.}
\end{figure*}

In our investigation, we followed the method of \cite{2013_McSween_a} and \cite{2013_Ammannito_a} while using the BI, SF1, and SF2 bands. Band centers are calculated following the method described in Sect. \ref{Subsec_spectral_parameters} and Appendix \ref{Appendix_BC_method} for both VIR and HED meteorite data. Maps of the BI, SF1, and SF2 band centers are reported in Figs. \ref{Fig_Appendix_BI_map}-\ref{Fig_Appendix_SF2_map}, Appendix \ref{Appendix_maps_band_center}. HED data come from the RELAB database\footnote{Reflectance Experiment Laboratory: \url{http://www.planetary.brown.edu/relab/}} and a selection is applied to keep the best-quality samples (140 in total, see Appendix \ref{Appendix_HED}); the VIR data are also filtered to discard noisy data (see Appendix \ref{Appendix_VIR}). We plot the SF1 center versus the BI center of the VIR and HED data, as shown in Fig. \ref{Fig_Scatterplot_SF1_BI}. Subsets of VIR data are defined following the HED distribution: we attributed different lithologies to each of them, labeling them from $1$ (eucrite) to $5$ (diogenite) according to the HED type to which they belong (represented by red, green and blue boxes in Fig. \ref{Fig_Scatterplot_SF1_BI}). Data that correspond to howardite and diogenite or to howardite and eucrite are labeled $4$ and $2$, respectively; those for which no lithology is assigned are excluded. Based on this classification, we obtain the map in Fig. \ref{FIG_Lithology_1_map}, which represents the qualitative lithology on the surface of Vesta. To limit the noise and make the map and its associated color scale continuous, an average is calculated for the data that overlap each other. In Table \ref{Table3}, we report the mean band centers of the VIR data falling within each class.\par
\begin{figure}
    \includegraphics{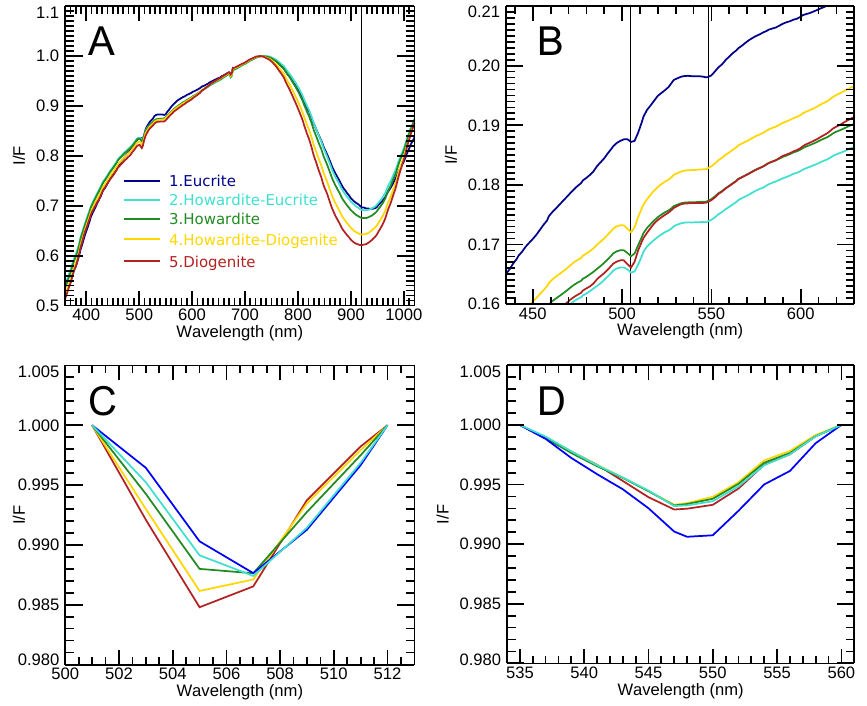}
    \caption{\label{FIG_Lithology_2_spectra}Spectra based on the Howardite--Eucrite--Diogenite classification of Fig. \ref{Fig_Scatterplot_SF1_BI}. (a) Median spectra normalized at \unit{730}{\nano\meter}. (b) Zoom-in on the \unit{440-630}{\nano\meter} range. (c) and (d) Continuum- removed SF1 and SF2 bands.}
\end{figure}

Figure \ref{FIG_Lithology_2_spectra} shows the spectra of the main end-members based on the eucrite--diogenite classification (boxes) of Fig. \ref{Fig_Scatterplot_SF1_BI}. For BI and SF1, panels A to C illustrate the shift towards shorter wavelengths as the lithology evolves from the eucrite to the diogenite type. For BI, the shift is as high as \unit{15.9}{\nano\meter} (from \unit{920.1}{\nano\meter} to \unit{936.0}{\nano\meter)} for the diogenite to eucrite classes, respectively. However, we note an important gap between the eucrite and the eucrite--howardite classes (\unit{8.2}{\nano\meter}). For SF1, the shift is small (about \unit{1}{\nano\meter}) but increases constantly from diogenite to eucrite (Table \ref{Table3}). In addition, the profile of the band distinctly evolves (panels B and C of Fig. \ref{FIG_Lithology_2_spectra}), although we reach the VIR resolution limit. 
This is not the case for SF2: the map of its band center (Fig. \ref{Fig_Appendix_SF2_map}) and panel D of Fig. \ref{FIG_Lithology_2_spectra} emphasize that no meaningful shift of the band center or change in the profile are observed (see below). A similar behavior was reported by \cite{2007_Klima}, who showed that the SF2 band centers of synthetic pyroxenes do not significantly shift as the iron content evolves (unlike the SF1 band). Here we confirm that no lithological information can be derived at Vesta using the SF2 band center.\par
We note that pure eucrite and diogenite end-members are poorly represented in VIR data (Fig. \ref{Fig_Scatterplot_SF1_BI}). Although this may partly stem from the prevalence of howardite on Vesta, it could also be ascribed to the different spectral resolution between the laboratory and the VIR spectra. This phenomenon could also be attributed to the spatial resolution of the VIR observations; that is, as several end-members are observed together in a single VIR pixel, the outcome tends to be more howarditic on average.\par
The map based on the SF1--BI couple (Fig. \ref{FIG_Lithology_1_map}) follows the same trend as the one presented in \cite{2013_Ammannito_a}, although the eucritic type seems to be less represented in Fig. \ref{FIG_Lithology_1_map}. The mineralogical meaning  of
the derived lithology and its implications for the evolution of Vesta   have been discussed by \cite{2013_Ammannito_a}; here we described the present map of the lithology in light of the maps of Sect. \ref{Sec_Maps}. Surface features are barely visible in Fig. \ref{FIG_Lithology_1_map}, unlike in the spectral parameter maps. For instance, Oppia (21) and Octavia (10) ejecta and the associated orange patches \citep{2013_LeCorre} cannot be distinguished in Fig. \ref{FIG_Lithology_1_map} while they clearly stand out on the color composite and spectral slope maps (Figs. \ref{Fig_RGB_380_465_730} to \ref{Fig_Slope_465_730}). Also, Aricia Tholus crater (11) and its counterparts (e.g., Occia (2)), that is, impacts that brought dark material to Vesta's surface, are not particularly distinct in the HED map. This is likely due to the fact that the spectral variability in those areas ---emerging only in the color composites and spectral slope maps--- is dominated by the variation in their nonpyroxenic material fraction.\par 
In general, we do not observe places that are purely eucrite-rich. In the southern hemisphere, only four craters stand out from the background. The strongest signature is represented by Tuccia (13) crater. Others signatures are associated to an unnamed crater at $12\degr$E--$38\degr$S; closer to the south pole, the western rim of Tarpeia (3) crater; and the rim of another unnamed (probably young based on the FC map) crater at $127\degr$E--$67\degr$S (north of Severina (9) crater). Thus, the eucritic end-member is mostly spatially diffuse on the surface, mixed with a howardite-like lithology, and is particularly distributed toward the west of Marcia crater (12) and to the southwest of Octavia (10). In the two cases, the eucritic and the howarditic lithologies are not associated with specific geologic units \citep{2014_Williams_c} and so likely correspond to the background unit instead of being excavated by impact craters.\par
Although the diogenite type is also poorly represented, it is easier to identify rich spots on the surface of Vesta. Such diogenite-rich spots are much more localized than the eucrite-rich areas and correspond to Matronalia Rupes (6), Lollia (7) crater (north to Matronalia Rupes (6)), an unnamed crater at $\sim120\degr$E--$23\degr$S (south to Serena (8) crater), the northern rim of Mamilia crater ($290\degr$E--$49\degr$N, not fully visible on the map), or the rims of Justina (22) and Severina (9) craters. In the latter case, a crater within Severina (9) could have exposed a larger amount of diogenite by excavating material below the floor of Severina (9). Antonia crater (14) and its surroundings are enriched in diogenite as well, but the complex geologic context \citep{2014_Kneissl, 2015_Zambon} and the limited resolution of the map in Fig. \ref{FIG_Lithology_1_map} make it difficult to attribute the diogenite-rich spots to different ejecta layers. Between $23$--$82\degr$E and $15$--$30\degr$N, other features show a diogenite signature. These remain localized and seem to correspond to the ejecta and rims of small young craters (based on the FC map), just like for Rubria (1) and Occia (2) craters. The diogenite-rich areas are generally linked to green color on the RGB Clementine composite and to high HBA (Figs. \ref{Fig_Clementine} and \ref{Fig_Half_BA}, respectively), although some exceptions can be seen, as in Tuccia (13), Canuleia (19), and Charito (20) craters.\par
The howarditic material is partly distributed in Vestalia Terra (between Marcia (12) and Oppia (21) craters) where it tends to be enriched in eucrite. Additionally, howarditic material is located between $\sim0\degr$E and $\sim90\degr$E where it is rather enriched in diogenite, like the Rheasilvia basin. This wide band is visible in some spectral parameter maps of Sect. \ref{Sec_Maps} and hypothesized as Rheasilvia crater ejecta \citep{2012_Reddy_a,2013_Ammannito_a}.\par
The map in Fig. \ref{FIG_Lithology_1_map} completes the previous map established by \cite{2013_Ammannito_a} and the distribution of the observed lithology confirms their results. The localized diogenite-rich spots are mainly found in the vicinity of Rheasilvia basin and were probably exposed by Rheasilvia and subsequent impact craters. The scarcity of diogenite surface patches favors the existence of a diogenite-rich lower crust and the magma ocean model that propose such a structure. Such a lower crust is exposed on the surface only where large impacts were able to excavate the upper crust (as in the south pole basins), exposing the diogenite-rich lithology. Other diogenite patches are found far from the south pole (e.g., Mamilia) and their origin remains unknown; as they are found in areas of relatively low altitude, this could mean that the diogenite-rich layer is present near the subsurface or the diogenite patches are ejecta from the south pole basins. An alternative explanation is the presence of diogenite-rich plutons under the surface that could be sampled by impacts \citep{2017_Raymond}. However, the small scale of the patches and their associated craters makes this hypothesis difficult to verify with our dataset.\par
\begin{table}
\caption{Band centers from the Fig. \ref{FIG_Lithology_2_spectra}}
\label{Table3}
\centering
\begin{tabular}{c c c}
\hline\hline
\multirow{2}{*}{\begin{tabular}[c]{@{}c@{}}\#-Lithology\end{tabular}} & \multicolumn{2}{c}{Band center (nm)}\\ \cline{2-3} & BI & SF1\\
\hline
1-E & $936.0$ & $506.9$
\\[1pt]
2-HE & $927.8$ & $506.6$
\\[1pt]
3-H & $925.7$ & $506.3$
\\[1pt]
4-HD & $922.5$ & $506.0$
\\[1pt]
5-D & $920.1$ & $505.9$
\\[1pt]
\end{tabular}
\tablefoot{VIR band centers calculated using the data of each class of Fig. \ref{Fig_Scatterplot_SF1_BI}. These also correspond to the spectra of Fig. \ref{FIG_Lithology_2_spectra}.}
\end{table}
%
\subsection{The key role of  impact craters in the surface spectral diversity}
\label{SubSec_impact's_impacts}
The spectral parameter maps show that impact craters play an important role in driving the surface spectral variation. We first emphasize that they are responsible for the broad low-albedo area that is highly visible in Fig. \ref{Fig_IoF_550} between $45\degr$S--$45\degr$N and $60$--$220\degr$E, through the delivery of chondritic-like material \citep{2012_McCord,2012_Reddy_b,2012_Prettyman,2012_De_Sanctis_b}. The clearer crater of this population ---probably because it is the youngest--- is on top of Aricia Tholus (11). These impacts brought carbonaceous material at the relatively small crater scale but the subsequent impacts spread this dark agent over time \citep{2012_McCord}. In the spectral parameter maps, this area is also visible through the R2 and R3 color composites (Figs. \ref{Fig_Ratio_550} and \ref{Fig_Clementine}), appearing blue and purple, respectively. In $S_{380-465nm}$, we note a lower slope, which is in favor of a deeper \ce{Fe-O} absorption. This would be in agreement with GRaND measurements that show a positive correlation between this low-albedo area and the \ce{Fe} enrichment there \citep{2013_Yamashita}. The low-albedo area is also correlated with the HBA (Fig. \ref{Fig_Half_BA}), which is lower in this region. \cite{2012_De_Sanctis_a} and \cite{2015_Frigeri_a} also report a relatively low BI depth there. Finally, the $S_{465-730nm}$ does not provide evidence for an albedo dichotomy of Vesta at a large scale. However, the spatial resolution achieved with our maps could be insufficient to reveal a $S_{465-730nm}$ trend at the scale of the small dark features mapped by \cite{2012_McCord} and \cite{2012_Reddy_b} with the Framing Camera.\par
Other craters generally reflect the composition of the crust. The color diversity reported on the color composite maps among the crater population support this point. For instance, Arruntia (5), Octavia (10), Vibidia (16), Teia (18), Cornelia (15), Oppia (21), and Justina (22) craters are nearly all unique in terms of colors through the color composites and are partly different in the HBA or the spectral slope maps. The giant impacts that formed the Veneneia and Rheasilvia basins are probably the most significant events that Vesta has undergone. These left their signature on the south pole, for example through diogenite-rich spots associate with craters or the Matronalia Rupes (6) scarp. These impacts also spread ejecta across the surface of Vesta, notably between $0\degr$E and $\sim60-90\degr$E as emphasized by some of the spectral parameters.\par
\subsection{The olivine-rich spots}
\label{SubSec_olivine}
Olivine was expected to be present at Vesta but its abundance was overestimated during the pre-Dawn era (e.g., \cite{1997_Gaffey,1997_Binzel}). In addition, in a global magma model, olivine should be preferably detected in the vicinity of the Rheasilvia basin, because the impact would have excavated olivine-rich mantle material. On the contrary, Dawn investigations revealed only a dozen olivine-rich locations across the surface; some of them are probably linked to Rheasilvia ejecta \citep{2014_Ruesch_a} but the main ones, at Bellicia (4) and Arruntia (5) craters, are clearly disconnected from the impact basin and located far from it, in the northern hemisphere \citep{2013_Ammannito_b,2014_Thangjam}. Despite the challenge of identification of olivine in pyroxene mixtures \citep{2013_Beck} or the questions surrounding the aforementioned detection \citep{2015_Combe_b}, it would be worthwhile investigating whether or not features attributed to olivine-rich spots show peculiar behavior on the spectral parameter maps presented in this study.\par
Among the maps of Sect. \ref{Sec_Maps}, only the R1 and the R2 color composites (Figs. \ref{Fig_RGB_380_465_730} and \ref{Fig_Ratio_550}) show distinct colors from the rest of the surface, namely a very diffuse light pink and a purple color, respectively. These variations are observed where olivine spots are detected \citep{2013_Ammannito_b, 2014_Thangjam}: in the northeast rim of Bellicia (4) and in the south to northwestern closest ejecta of Arruntia (5). In the latter, the purple color observed in the R2 map (Fig. \ref{Fig_Ratio_550}) must be distinguished from the red color (corresponding also to high $S_{465-730nm}$ values) that characterizes most of the ejecta of Arruntia (5). The olivine-rich spots have been incorrectly assigned to eucritic-rich areas in the lithological map of Fig. \ref{FIG_Lithology_1_map} because the BI center shifts toward longer wavelengths, which is due to the asymmetric shape of BI caused by olivine itself \citep{1981_Singer,1986_Cloutis} (see also Fig. \ref{Fig_Appendix_BI_map}). We do not observe other peculiar characteristics in the other maps of the spectral parameters. The spatial resolution of our maps also prevents us from identifying the smallest olivine-rich craters around Bellicia (4) \citep{2013_Ammannito_b} or the various spots reported by \cite{2014_Ruesch_a} across the surface.
%
\section{Conclusions}
\label{Conclusion}
The newly corrected and calibrated dataset of the VIR visible channel allow us to study the surface of Vesta  in greater detail. The combination of color composites, spectral slopes, and band parameters allows us to highlight similarities and differences among the surface features of Vesta thanks to the almost global coverage.\par
Our spectral parameters show an important spectral diversity on the surface that is highlighted by (and is caused by) impact cratering. Indeed, the observed spectral diversity, as revealed by craters, indicates that the subsurface of Vesta is not homogeneous and the regions more deeply excavated, as in the south polar basins, show an enrichment in diogenite. Moreover, some impacts, by bringing exogenous dark material, also contribute to modifying the spectral properties of Vesta, primarily by darkening and hydrating large parts of its surface.\par
The pyroxene-rich surface is characterized by two prominent bands located at \unit{0.9}{\micro\meter} and \unit{1.9}{\micro\meter}. Infrared and visible data are generally used in combination with other techniques to study the HED-related lithology. The use of the visible wavelengths alone can be a limiting factor in that case. In our study, thanks to the high spectral resolution of VIR, we identified the small spin-forbidden bands at \unit{506}{\nano\meter,} and those at \unit{550}{\nano\meter} for  the  first  time. By monitoring the center shift of the \unit{930}{\nano\meter} and \unit{506}{\nano\meter} bands, we successfully retrieved the lithology of Vesta at a nearly global scale. Our results agree with the lithology previously derived by for example \cite{2013_Ammannito_a} and thus demonstrate that visible data alone are sufficient to obtain advanced mineralogical information. \cite{2007_Klima} did not observe a significant center shift in the \unit{550}{\nano\meter} band in synthetic pyroxenes. We noted a similar behavior across the surface of Vesta, which prevented us from using the \unit{550}{\nano\meter} band to derive lithological indications.\par
Olivine-rich spots on Vesta, as identified by \cite{2013_Ammannito_b} and \cite{2014_Thangjam}, show a distinct color in some of our spectral parameter maps (R1 and R2 color composites, Figs. \ref{Fig_RGB_380_465_730} and \ref{Fig_Ratio_550}) although this color behavior is not a direct diagnostic. The limited spatial resolution prevents other olivine-rich spots from being identified.\par

%
\begin{acknowledgements}
VIR is funded by the Italian Space Agency (ASI) and was developed under the leadership of INAF-Istituto di Astrofisica e Planetologia Spaziali, Rome, Italy (ASI-INAF grant I/004/12/0). The instrument was built by Selex-Galileo, Florence, Italy. The authors acknowledge the support of the Dawn Science, Instrument, and Operations Teams.
The authors made use of TOPCAT (Tools for OPerations on Catalogues And Tables, \cite{2005_Taylor}) for a part of the data analysis and figure production. This research has made use of "Aladin Desktop" developed at CDS, Strasbourg Observatory, France \citep{2000_Bonnarel, 2015_Fernique}. This research utilizes spectra from the NASA RELAB facility at Brown University. We appreciate helpful discussions with J. Brossier (INAF-IAPS) on manuscript editing. We thank S. Schröder for his constructive review that improves the quality of the manuscript.
\end{acknowledgements}

\bibliographystyle{aa.bst}
\bibliography{main.bib}
\begin{appendix}
\section{Density maps}
\label{Appendix_Density_maps}
\begin{figure*}
    \centering
    \includegraphics[width=17cm]{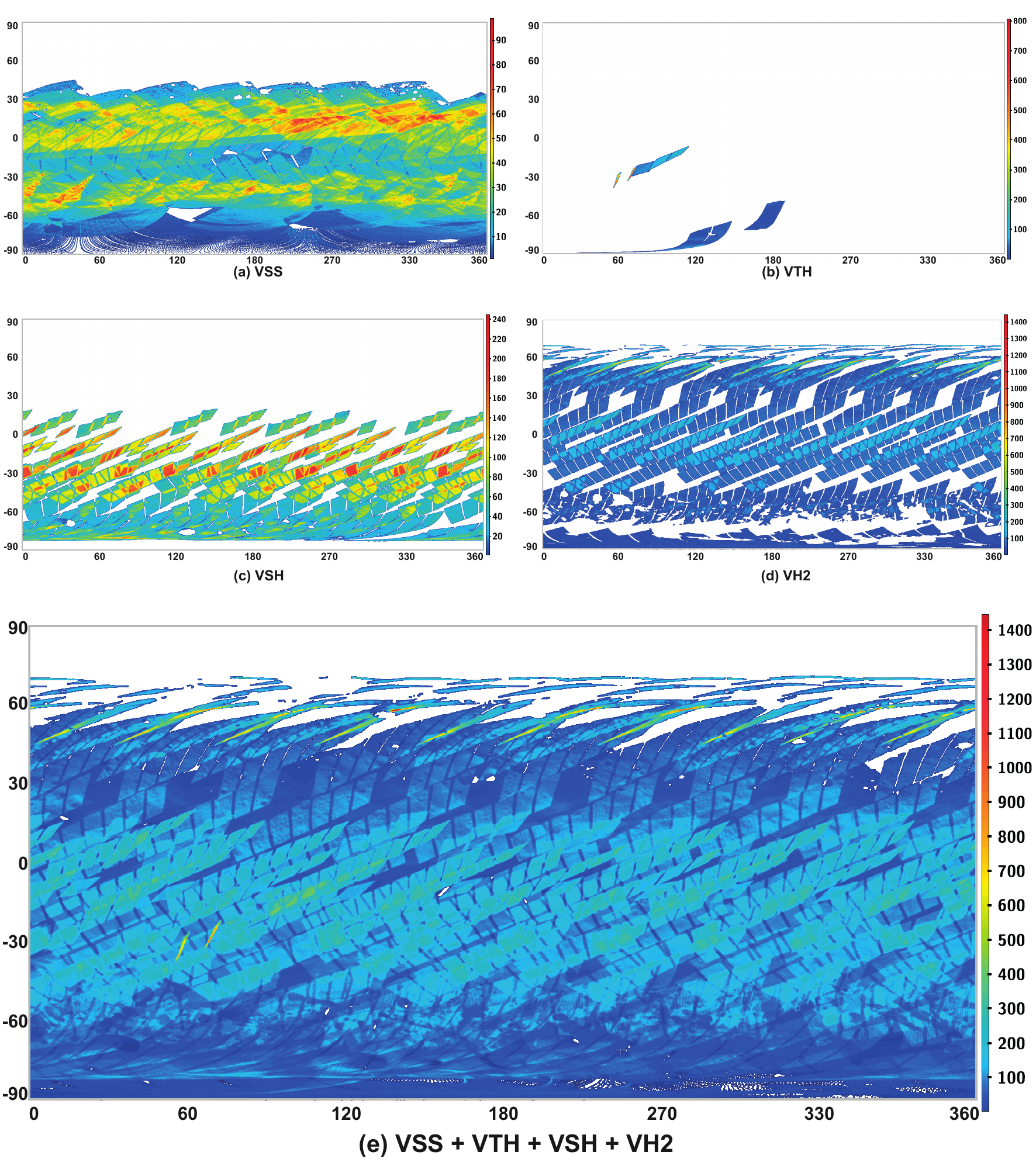}
    \label{Fig_Appendix_DENSITY_MAP}
    \caption{Density maps of the VIR visible data set used in the study. Panel A corresponds to the VSS mission phase; panel B to the VTH; panel C to VSH; panel D to VH2; and panel E regroups every four. For details about the mission phases, see Table \ref{Table1} and Sect. \ref{Subsec_Data_corr}. Each map is built with TOPCAT with a Plate Carrée projection (see Sect. \ref{Subsec_maps_proj}), and observations are represented as points. The scale corresponds to the square root of the observation density.}
\end{figure*}
\section{Band centers - Method}
\label{Appendix_BC_method}
\begin{figure}
    \centering
    \includegraphics{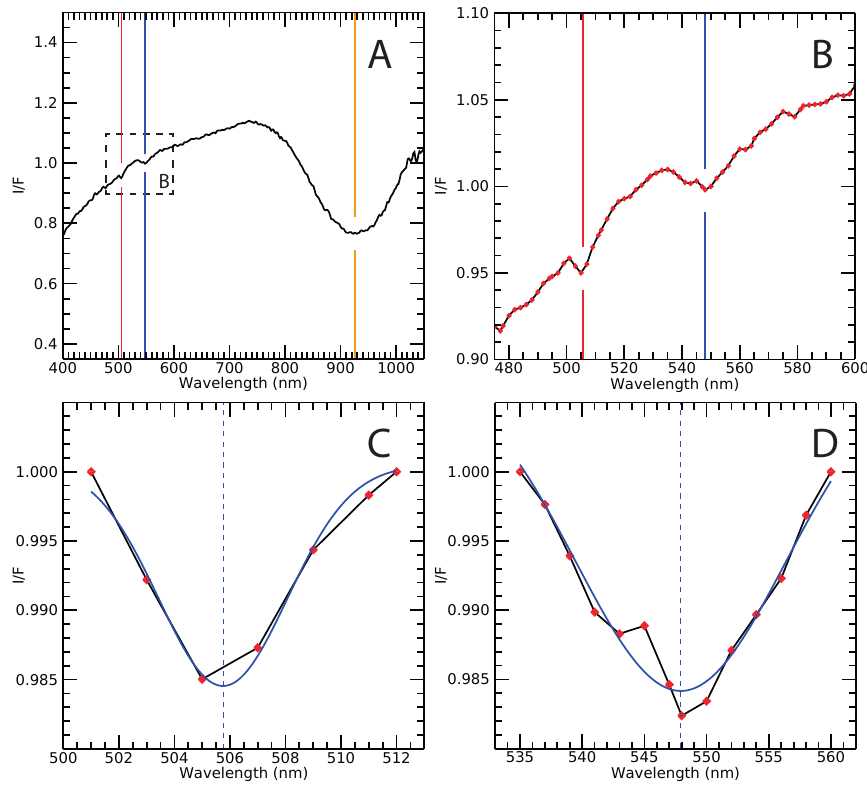}
    \caption{\label{Fig_Appendix_SF_band_fit} Example of a single VIR visible spectrum and results of the SF1 and SF2 fits. a) VIR visible spectrum from \unit{400}{\nano\meter} to \unit{1050}{\nano\meter}. b) Zoom-in on the SF1 and SF2 bands between \unit{475}{\nano\meter} to \unit{600}{\nano\meter}. c) and d) Continuum removed SF1 and SF2 bands, respectively. In the different panels: the red, blue and orange plain vertical lines indicate the center of the SF1, SF2 and BI, respectively. Red symbols correspond to the original positions of the VIR spectral channels. The plain blue curves show the results of the fits by the Gaussian functions recomputed with a sampling of \unit{0.1}{\nano\meter}. The dashed vertical blue lines indicate the peak centroids of the Gaussian functions, corresponding to the band centers.}
\end{figure}

The SF1 and SF2 band centers are fitted using a Gaussian function made of four free parameters (peak value, peak centroid, gaussian sigma and a constant baseline). We use the MPFITPEAK routine for IDL (Interactive Data Language). The use of a constant baseline allows us to fit the continuum removed spectrum that has a non-null baseline. In addition, if one or both wings of the fitted band are affected by the noise, having a non-null constant baseline provides more flexibility to fit the band. The peak centroid is defined as the band center as illustrated by the Fig. \ref{Fig_Appendix_SF_band_fit}.\par
We note that in many cases the noise of VIR spectra is too high at this spectral resolution (few nanometers), leading the fit to fail or to return anomalous results. However, as detailed in Sect. \ref{Discussion} and Appendix \ref{Appendix_VIR}, we filtered the data when necessary (i.e., for the study of the lithology) while the high redundancy of VIR observations allows a statistically meaningful result.
\section{Framing Camera HAMO map and main features}
\label{Appendix_HAMO}
\begin{figure*}
    \centering
    \includegraphics[width=17cm]{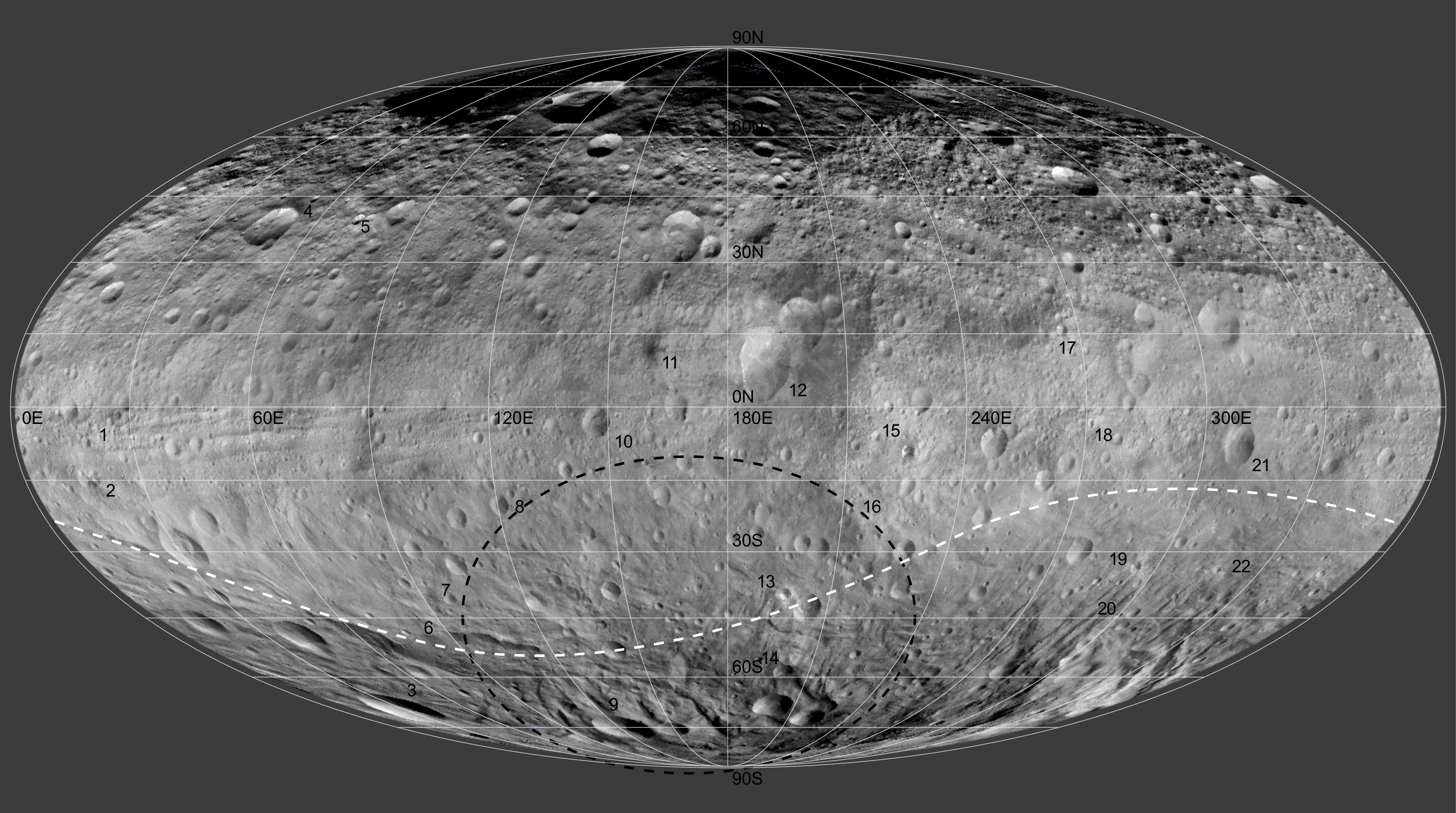}
    \label{Fig_Appendix_HAMO_MAP}
    \caption{Framing Camera HAMO map from \cite{2015_Roatsch} reprocessed as a HiPS (Hierarchical Progressive Surveys) and with a Mollweide projection. The map is used as background context for the maps of the spectral slopes in Figs. \ref{Fig_Slope_380_465}, \ref{Fig_Appendix_SLOPE_465_730}, and \ref{Fig_Half_BA}. Numbers refer to the features of Table \ref{Table2} discussed in the text. The white and dark dashed lines roughly delineate the Rheasilvia and Veneneia basins.}
\end{figure*}
\section{Maps of the spectral slopes without Framing Camera context and coordinate grid}
\label{Appendix_maps_no_transp}
\begin{figure*}
    \centering
    \includegraphics[width=17cm]{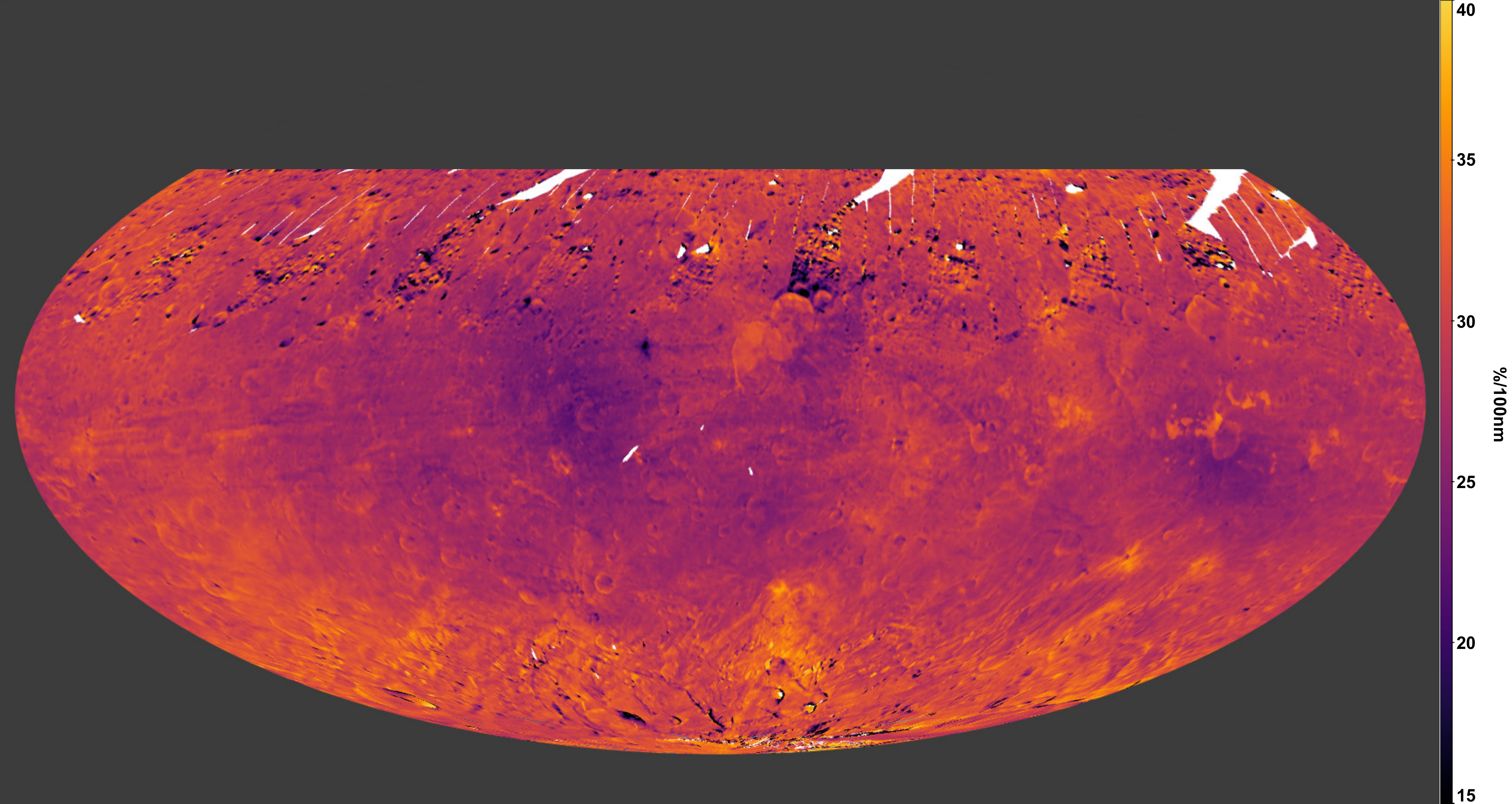}
    \caption{\label{Fig_Appendix_SLOPE_380_465} Map of the VIR $S_{380-465nm}$ spectral slope without transparency effect and Framing camera context. White areas correspond to missing data.}
\end{figure*}
\begin{figure*}
    \centering
    \includegraphics[width=17cm]{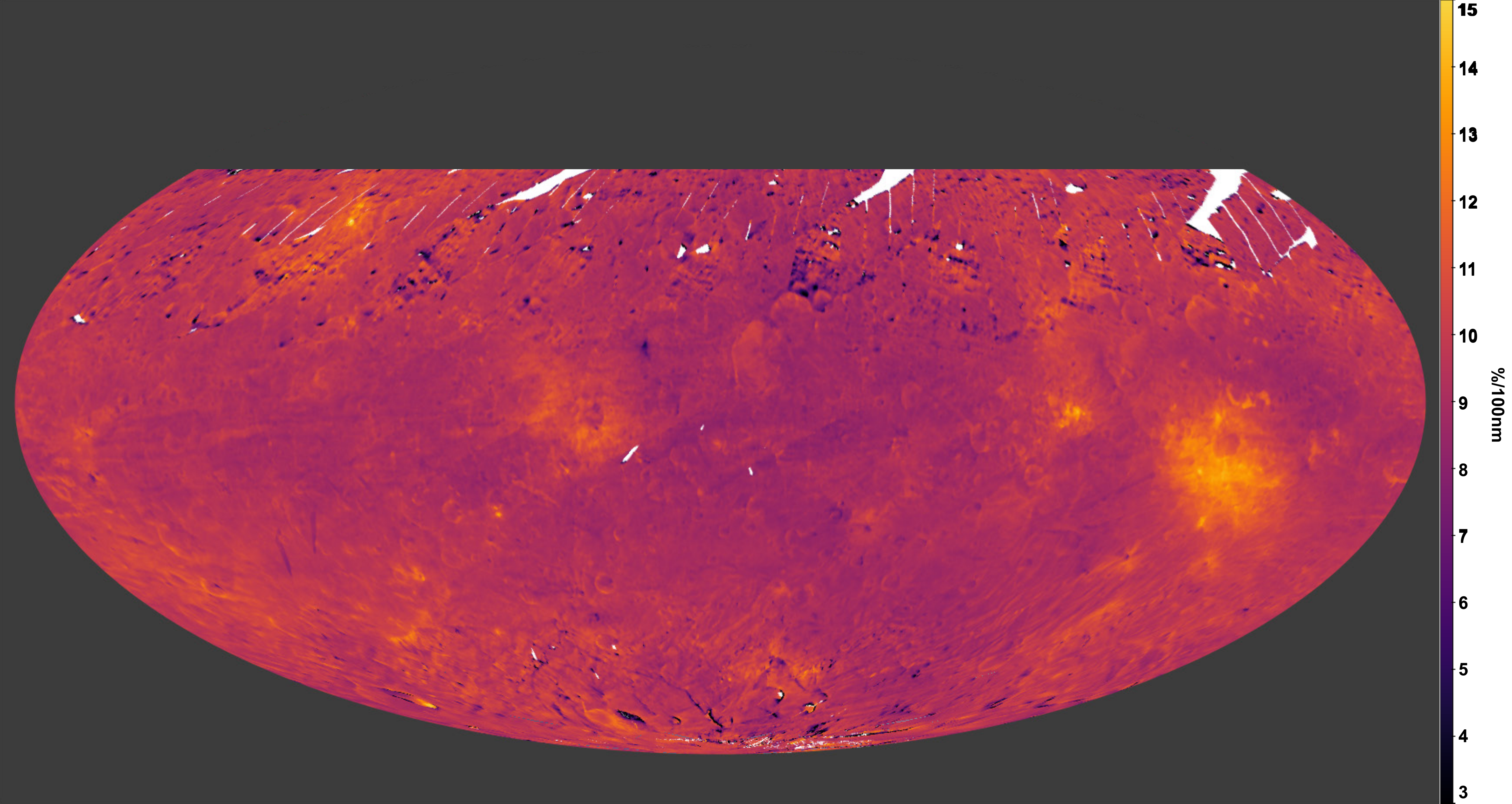}
    \caption{\label{Fig_Appendix_SLOPE_465_730} Same as Fig. \ref{Fig_Appendix_SLOPE_380_465} but for the VIR $S_{465-730nm}$ spectral slope.}
\end{figure*}
\begin{figure*}
    \centering
    \includegraphics[width=17cm]{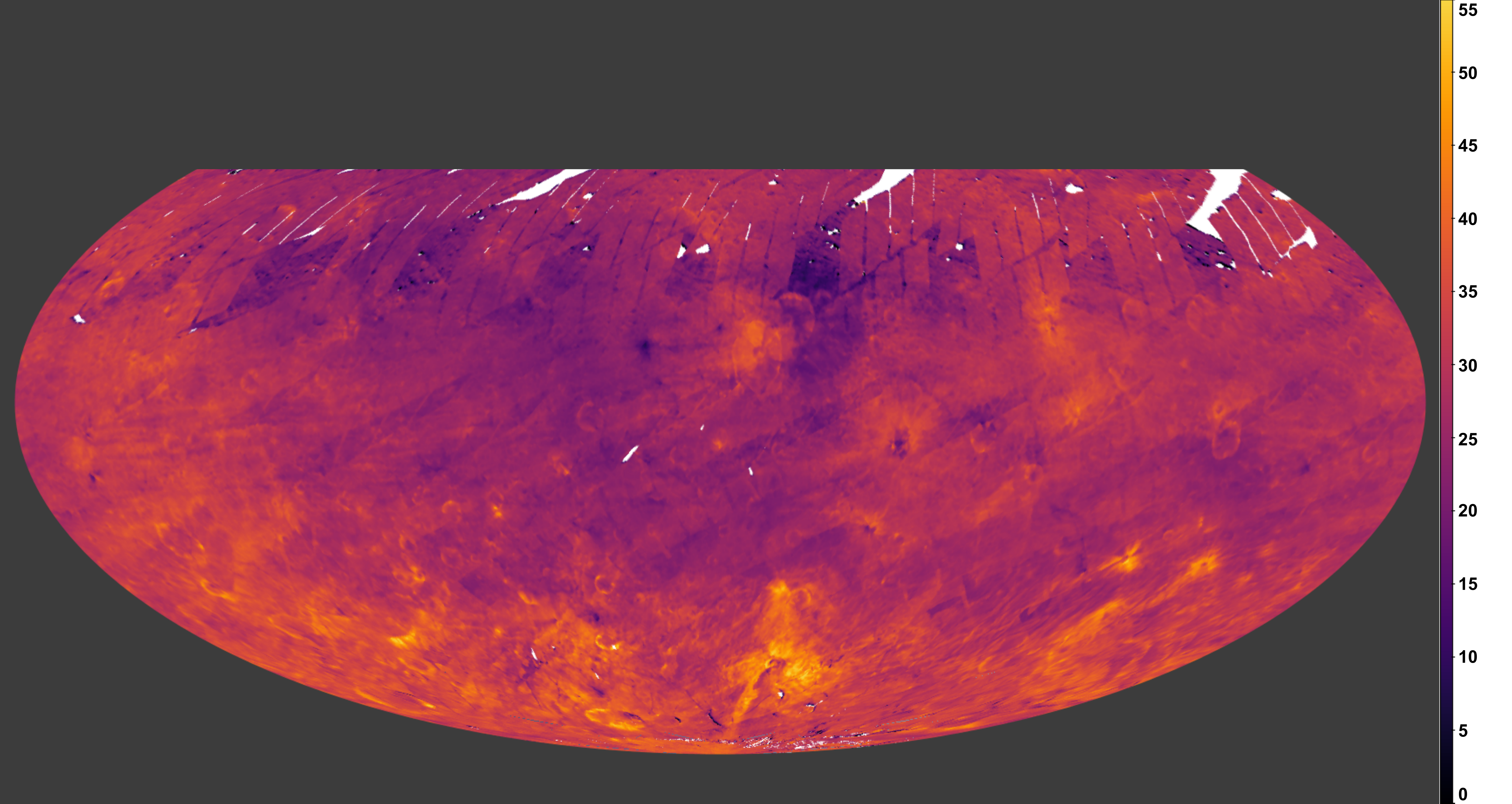}
    \caption{\label{Fig_Appendix_Half_BA} Same as Fig. \ref{Fig_Appendix_SLOPE_380_465} but for the VIR half band area.}
\end{figure*}
\section{Maps of the south pole}
\label{Appendix_South_pole}
\begin{figure*}
    \centering
    \includegraphics[width=17cm]{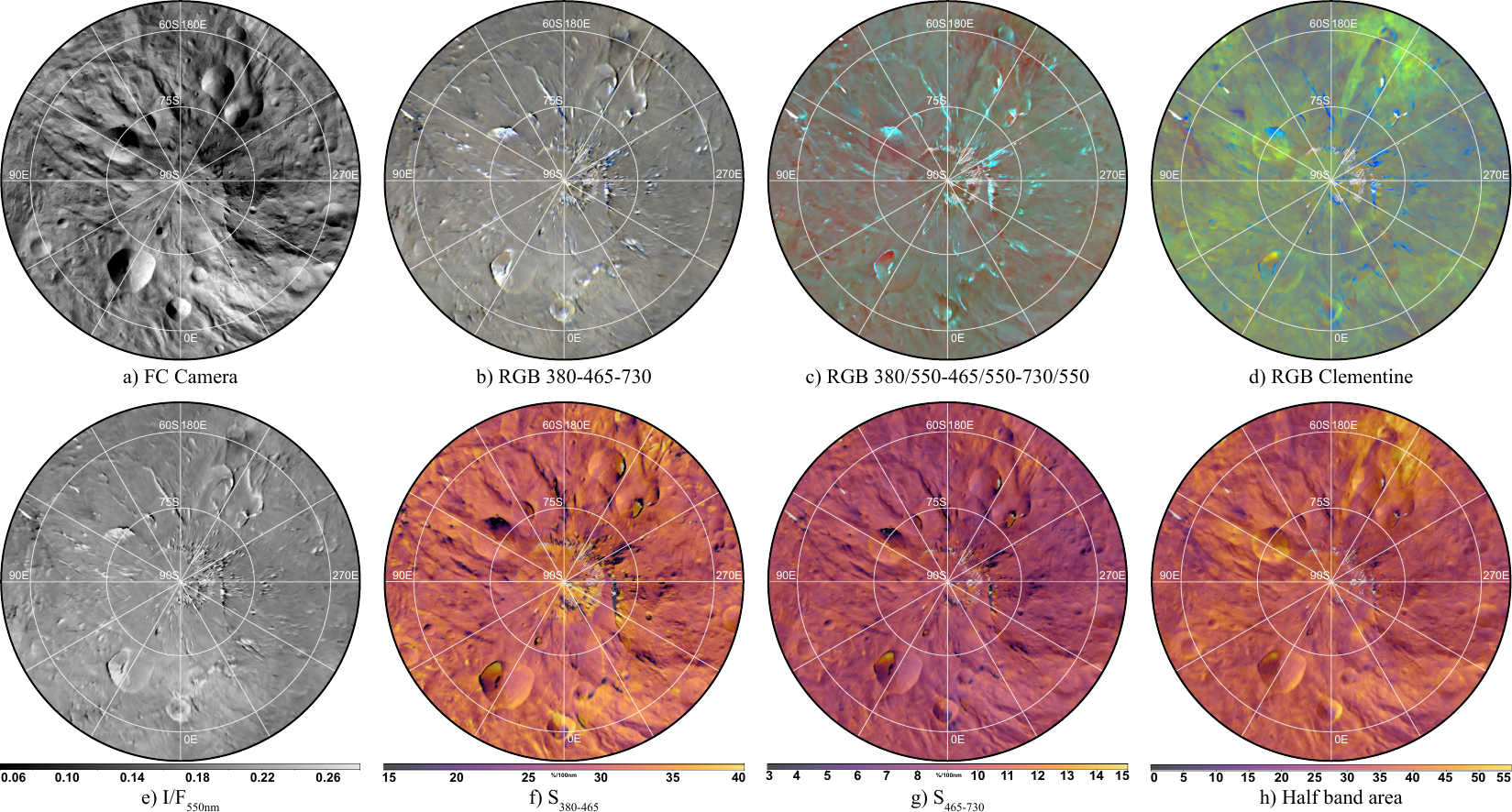}
    \caption{\label{Fig_Appendix_South_pole}Map of the south pole of Vesta. Numbers refer to the features of Table \ref{Table2} discussed in the text. The central peak fills the upper-left quarter of the image, from $90$ to $180\degr$E and from $0$ to $60\degr$S;}
\end{figure*}
\section{Lithology - Methods and supplementary materials}
\label{Appendix_Lithology_supp}
\subsection{Howardite - Eucrite - Diogenite data}
\label{Appendix_HED}
A total of $281$ HED spectra were initially selected in the RELAB spectral catalog based on the following criteria: (a) the \textit{subtype} must be howardite, eucrite, or diogenite; (b) only powders are retained whereas meteorite chips are excluded; (c) the acquisition range is between \unit{300}{\nano\meter} and \unit{2600}{\nano\meter} (others spectra were compatible but this choice made the computation easier); (d) the spectral sampling is equal to \unit{5.0}{\nano\meter}. Some duplicates were excluded but others were retained when several grain sizes of one sample exist, so no filtering based on the grain size was applied. In a second phase, about $141$ spectra were excluded because of a poor-quality fit of the SF1 and SF2 bands or because of the spectra quality itself.\par
At the end, a total of $140$ HED spectra were used, made up of $41$ howardites, $79$ eucrites, and $19$ diogenites. Table \ref{Table4} reports their sample names, types, and band centers (BI, SF1, and SF2). The reader is referred to the RELAB catalog for more details about each sample.
\subsection{VIR data selection}
\label{Appendix_VIR}
\begin{figure*}
    \centering
    \includegraphics[width=17cm]{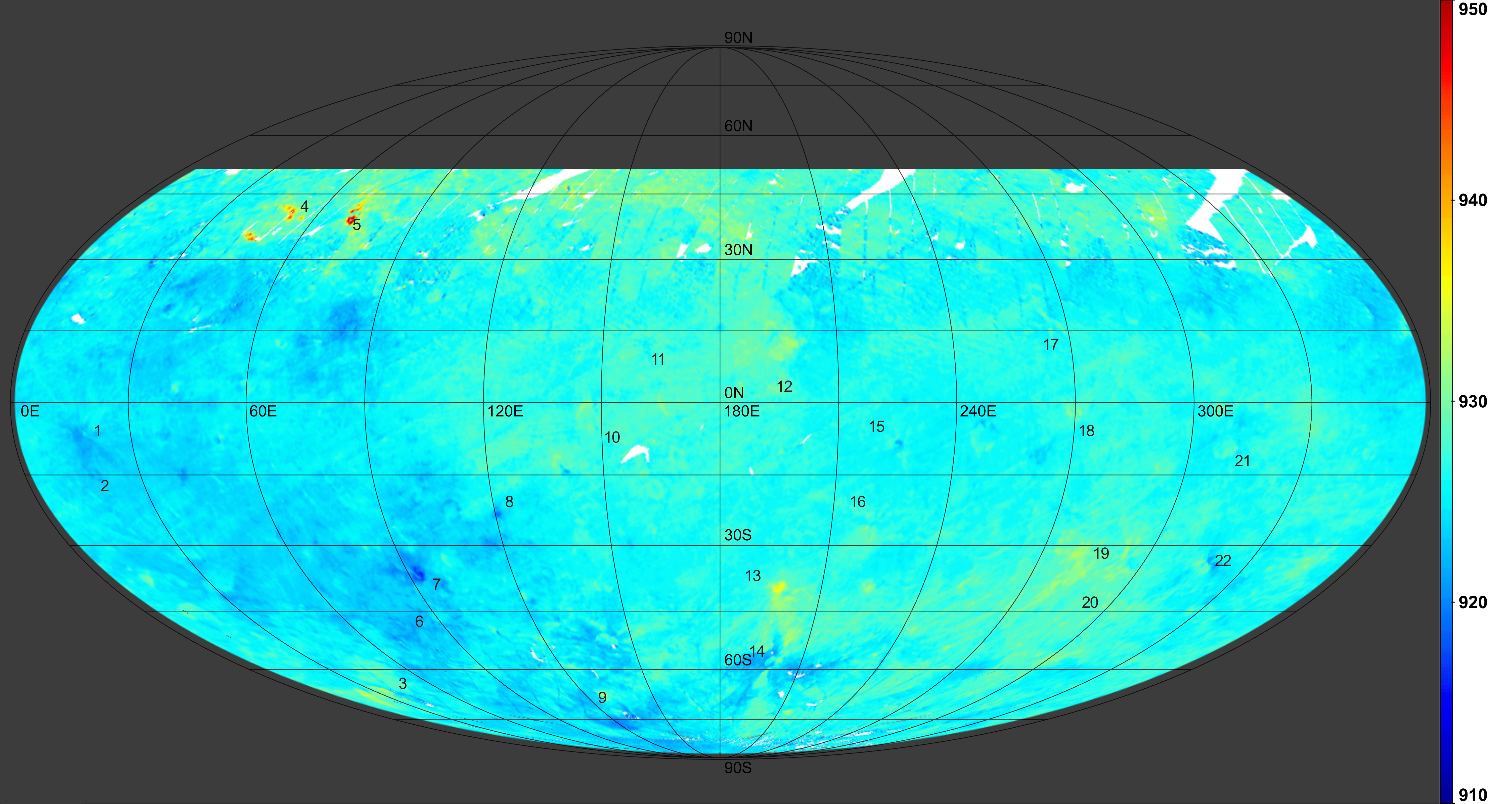}
    \caption{\label{Fig_Appendix_BI_map} Map of the BI band center.}
\end{figure*}

\begin{figure*}
    \centering
    \includegraphics[width=17cm]{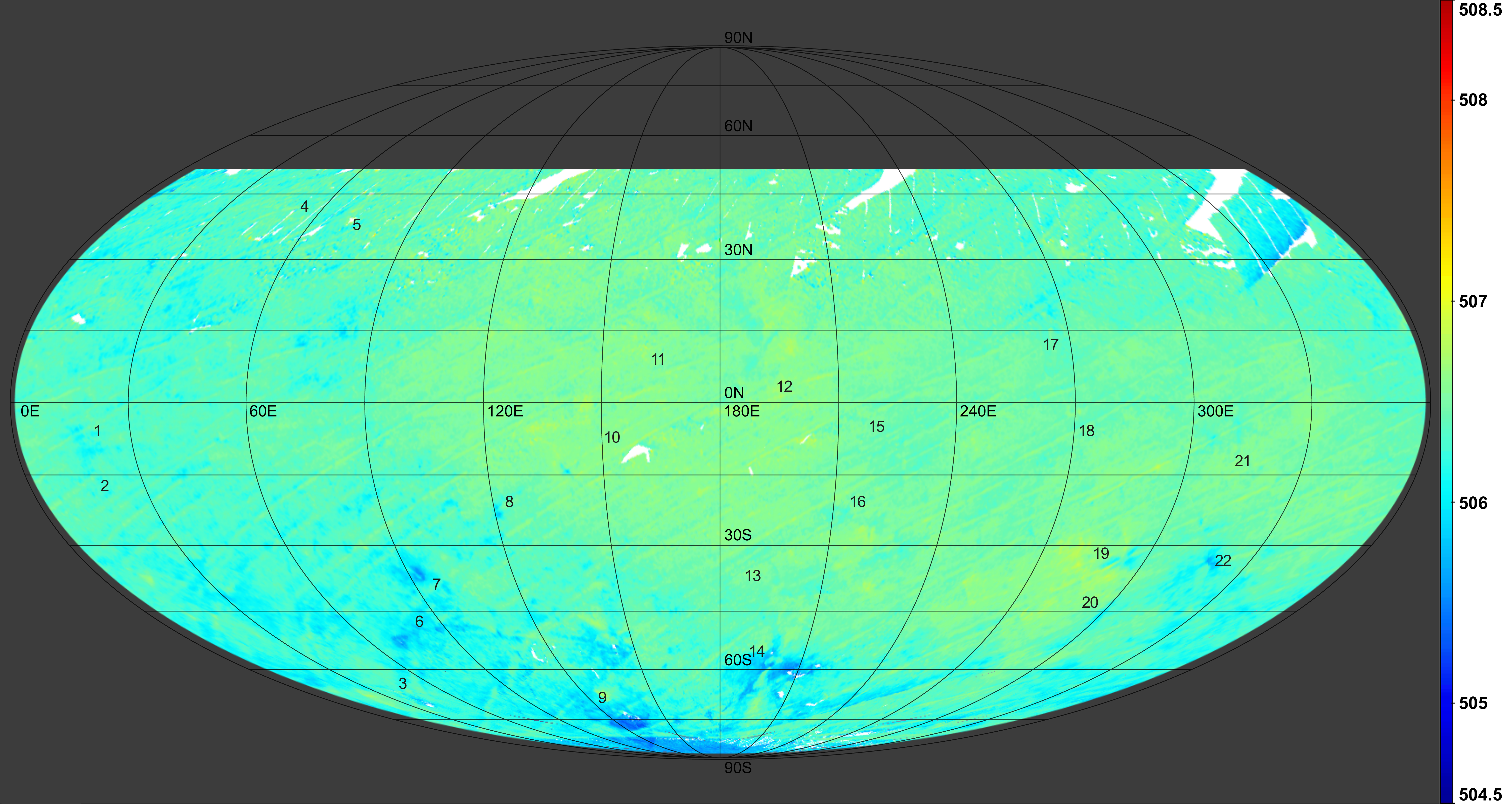}
    \caption{\label{Fig_Appendix_SF1_map} Map of the SF1 band center.}
\end{figure*}

\begin{figure*}
    \centering
    \includegraphics[width=17cm]{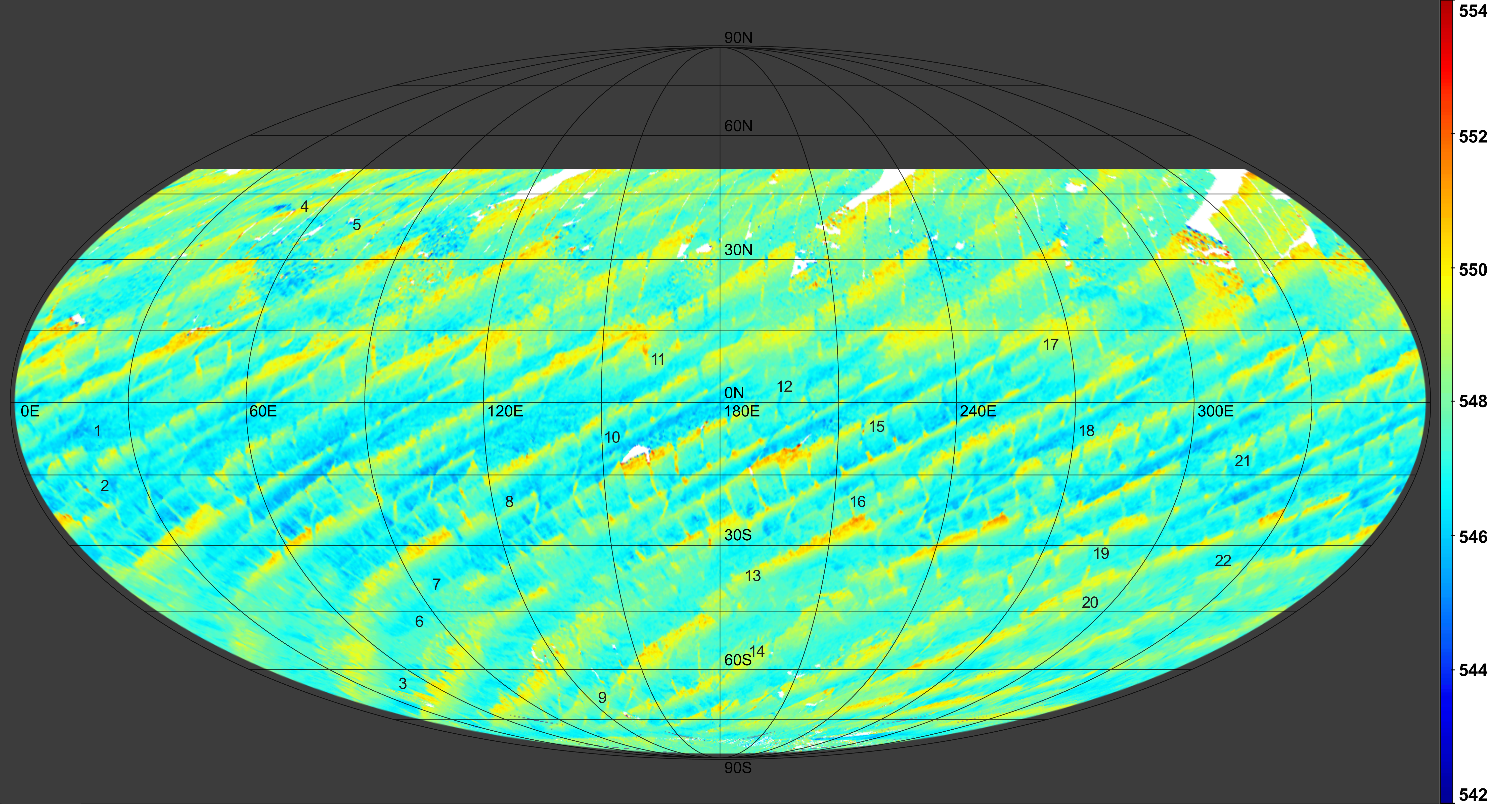}
    \caption{\label{Fig_Appendix_SF2_map} Map of the SF2 band center.}
\end{figure*}

Unlike the maps in Sect. \ref{Sec_Maps}, we exclude a part of the VIR data for the study of the lithology. This selection comes after the fitting procedure to determine the band centers. We excluded the spectra that do not satisfy the following criteria: (a) $1000$ DN at \unit{550}{\nano\meter}; (b) $\chi_{SF1}^2<2.10^{-4}$ and $\chi_{SF2}^2<1.10^{-3}$; (c) $\sigma_{SF1}>2$ and $\sigma_{SF2}>2$. Here, $\sigma$ refers to the Gaussian sigma. Points (b) and (c) refer to the parameters obtained from each fit. The filtering allows us to keep $78\%$ of the data: maps of the band centers (Figs. \ref{Fig_Appendix_BI_map}-\ref{Fig_Appendix_SF2_map}) use this dataset. About $3.5$ M spectra are discarded (of a total of $16$ M), corresponding mainly to noisy spectra or spectra for which the fit is not successful because of the noise for example. This drastically reduces the scattering of the VIR data distribution in Fig. \ref{Fig_Scatterplot_SF1_BI} but does not eliminate all the unwanted data. It is worthy to note that a more accurate selection is difficult to achieve and that filter thresholds have been defined empirically. Indeed, the fits of the SF1 and SF2 bands are very sensitive to the spectrum quality; those bands observed at the VIR spectral resolution limit. The advantage of the VIR data is the number of observations available that reinforce the final reliability of the process.\par
Through the classification process of the VIR data, following the HED band centers (see Sect. \ref{Subsec_Litho}), we also exclude the data that do not fall into a specific class (i.e., all the data out of the boxes of Figs. \ref{Fig_Scatterplot_SF1_BI}). After this classification $77\%$ of spectra remain (about $12.5$ M spectra) in the case of Fig. \ref{Fig_Scatterplot_SF1_BI} selection. The map of the lithology in Fig. \ref{FIG_Lithology_1_map} uses this dataset.
\subsection{Maps of the BI, SF1, and SF2 band center}
\label{Appendix_maps_band_center}
\includepdf[page={1-}]{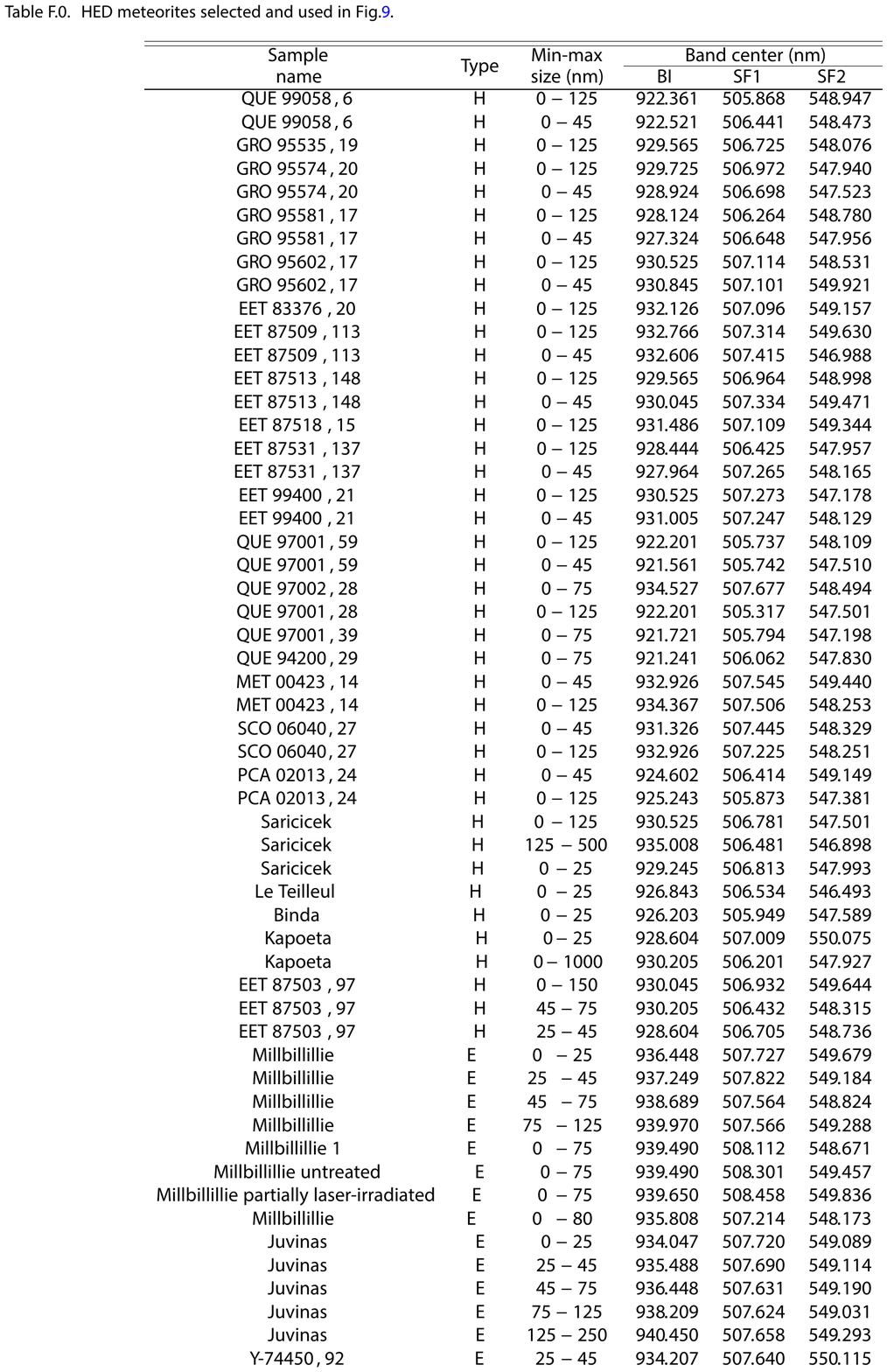}
\end{appendix}
\end{document}